\documentclass[journal]{IEEEtran}
\usepackage{cite}
\usepackage{amsmath}
\usepackage{graphicx}
\usepackage{amssymb}
\usepackage{mathdots}
\usepackage{arydshln}

\usepackage[para,online]{threeparttable}
\usepackage{enumitem}
\usepackage{stfloats}% <-- added
\DeclareMathOperator{\E}{\mathbb{E}}
\begin{document}
	\title{Stochastic Analysis of LMS Algorithm \\with Delayed Block Coefficient Adaptation}
	
	\author{Mohd.~Tasleem~Khan
		and Oscar~Gustafsson,~\IEEEmembership{Senior Member,~IEEE}% <-this % 
		\thanks{Received \today.}
		\thanks{The authors are with the Division of Computer Engineering, Department of Electrical Engineering, Link\"oping University, SE-581 83 Link\"oping, Sweden. E-mail: mohd.tasleem.khan@liu.se}
		\thanks{This work was supported by the ELLIIT strategic research environment.}
	}
	\maketitle
	\begin{abstract}

In high sample-rate applications of the least-mean-square (LMS) adaptive filtering algorithm, pipelining or/and block processing is required. As opposed to earlier work, pipelining and block processing are jointly considered to obtain what we refer to as the delayed block LMS (DBLMS) algorithm. Different stochastic analyses for the steady and transient states to estimate the step-size bound, adaptation accuracy, and adaptation speed based on the recursive relation of delayed block excess mean square error (MSE) are presented. The effect of different amounts of pipelining delays and block sizes on the adaptation accuracy and speed of the adaptive filter with different filter lengths and speed-ups are studied. It is concluded that for a constant speed-up, a large delay and small block size lead to a slower convergence rate compared to a small delay and large block size with almost the same steady-state MSE. Monte Carlo simulations indicate a good agreement with the proposed estimates for Gaussian inputs. 
	\end{abstract}
	\begin{IEEEkeywords}
		Convergence, Least-mean-square (LMS), Step-size, Stability, Mean-square error (MSE), DBLMS
\end{IEEEkeywords}
%\IEEEpeerreviewmaketitle
\section{Introduction}
%\usepackage{frcursive}
%\newenvironment{frcseries}{\fontfamily{frc}\selectfont}{}
%\newcommand{\textfrc}[1]{{\frcseries#1}}
%\newcommand{\mathfrc}[1]{\text{\textfrc{#1}}}
%\dots
%\mathbf{\mathfrc{P}}

\IEEEPARstart{A}{DAPTIVE} filtering algorithms are widely used in many signal processing applications such as system identification, noise/echo cancellation, channel equalization and linear prediction \cite{key-0, diniz1997adaptive, douglas1999introduction, sayed2003fundamentals,farhang2013adaptive}. For instance, in system identification, the transfer function of an unknown plant needs to be determined in order to apply the necessary control signals. In such scenarios, adaptive filters (ADFs) can be employed to model the unknown plant. To develop the ADFs, least-mean-square (LMS) algorithm is probably the most popular due to its simplicity, robustness, and ease of implementation \cite{liu2008kernel,bershad2008stochastic}. Nonetheless, it offers satisfactory convergence performance for a given choice of the step-size.
 Foreseeing anticipated demands of data rate in next decade technologies \cite{pinto2017compressed,khan2020high}, it is essential to develop high sample rate LMS ADFs using advanced signal processing algorithms \cite{sayed2003fundamentals, farhang2013adaptive, lima2017performance}. 

The sample rate of LMS ADFs is fundamentally limited by the iteration period bound caused by the coefficient adaptation loop. Increasing the sample rate can be achieved by modifying the LMS algorithm, typically in one of two different ways. The delayed LMS (DLMS) ADF \cite{long1989lms, meyer1990modular, long1992corrections, tobias2004leaky} introduces a set of $D$ additional delays in the adaptation loop, to decrease the iteration period, and therefore increase the sample rate by $D+1$. The block LMS (BLMS) ADF \cite{burrus1971block,clark1981block,feuer1985performance, lim1997optimum} processes a block of $L$ samples with an iteration period similar to the LMS algorithm. Hence, the sample rate is expected to be $L$ times higher. However, the increase in sample rates for these algorithms comes at the expense of slower convergence as the step-size bound is decreased. Note that these are first-order estimations of the sample rate, there are additional factors affecting it, but implementation results indicate that they are relevant \cite{khan2022asic}. 
  %\marginpar{The properties of these algorithms have not been examined as they are used to increase the sample rate. }
 
Efforts have been made to determine the step-size bounds of DLMS \cite{kabal1983stability,long1989lms,long1992corrections,rupp1994analysis,ernst1995analysis} and BLMS \cite{clark1981block, feuer1985performance, lee2012step}. For DLMS, Kabal \cite{kabal1983stability} estimated the step-size bound using the behaviour of the average coefficient error vector. Long \textit{et al.} provided an improved step-size bound using some simplifications \cite{long1989lms,long1992corrections}. 
In \cite{rupp1994analysis}, the step-size bound was obtained by transforming it into a problem of finding the eigenvalues of a matrix. However, the matrix size grows quadratically with $D$. Later, in \cite{ernst1995analysis}, an implicit step-size bound was derived in terms of a complex algebraic expression. 

In contrast to DLMS, the convergence performance of BLMS is similar to LMS, as long as the BLMS step size is $L$ times the LMS step size \cite{clark1981block}. Feuer \cite{feuer1985performance} argued that there were no guaranteed convergences of block mean square error (MSE) using the step size from \cite{clark1981block}. Therefore, an easy-to-calculate step-size bound was derived, more restrictive than \cite{clark1981block}. Using the framework from \cite{feuer1985performance}, Lee \textit{et al.} \cite{lee2012step} presented a not-so-tight bound for frequency-domain BLMS.   

\begin{figure}
	\centering
	\includegraphics[scale=0.65]{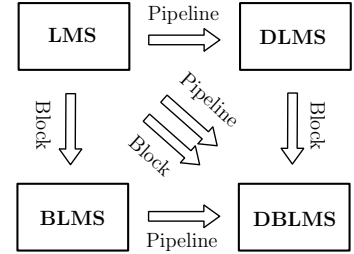}
	\caption{Relations between LMS, DLMS, BLMS, and the considered DBLMS algorithm via pipelining and/or block processing.}
	\label{fig_concept}
\end{figure}

From a high-speed implementation perspective, it is naturally of interest to consider both DLMS and BLMS. However, it is also possible to combine those two approaches into a delayed block LMS (DBLMS) algorithm. This is the focus of the current work. The relations between the different algorithms are illustrated in Fig.~\ref{fig_concept}. The DBLMS algorithm has a speedup of $S = (D+1)L$. It is clear that the DBLMS ADF reduces to LMS ADF for $D=0$ and $L=1$; DLMS ADF for $L=1$; and BLMS ADF for $D=0$.

In this paper, we jointly consider delayed block adaptive filtering procedures with the Wiener filtering problem based on the generalization of the gradient searching LMS algorithm of Widrow and Hoff \cite{widrow1971adaptive,widrow1976comparison,widrow1977stationary}. In the present work, we show that by introducing block into the DLMS algorithm (or delay into the BLMS algorithm), it is possible to trade the convergence properties of DLMS and BLMS together with a suitable combination of delay and block sizes, while ensuring algorithm stability. The important results here are design formulas for DBLMS ADFs. Cautiously, the designer has to consider the maximum usable delay and block size as it is limited by step size in a known way. The groundwork laid here is expected to be useful to develop efficient delayed and block procedures for other algorithms commonly found in estimation and detection theory. The key contributions of this paper are analytical first-order approximations for the following quantities expressed in $D$ and $L$, derived for DBLMS ADFs based on second-order statistics:
\begin{itemize}
	\item Bound on the adaptation step size, (\ref{eq:dblms_ss_ineq_2}), and the optimum adaptation step size in terms of MSE, (\ref{eq:opt_ss_small}).
	\item Excess MSE error, (\ref{eq:dblms_mse_excess_exp_ss}), misadjustment, (\ref{eq:misadj_general}), and misadjustment at optimum adaptation step size, (\ref{eq:misadjustment1}).
	\item Adaptation time constant, (\ref{eq:D8}), and slope, (\ref{eq:speed}).
\end{itemize}
Simulations are used to confirm the validity of the derived approximations.

The rest of the paper is organized as follows. In Section \ref{sec:lms}, the LMS algorithm followed by its delayed and block adaptation variants are reviewed. In the next Section, the delayed block LMS adaptive filtering procedure and algorithm are introduced. In Section IV, the mathematical analysis to obtain the measures for convergence properties of the DBLMS algorithm is performed. In Section V, the performance of the DBLMS algorithm with numerical simulations and analytical results is presented. Finally, conclusions are provided in Section VI.
\begin{figure*}
	\centering
	\includegraphics[width=1\linewidth]{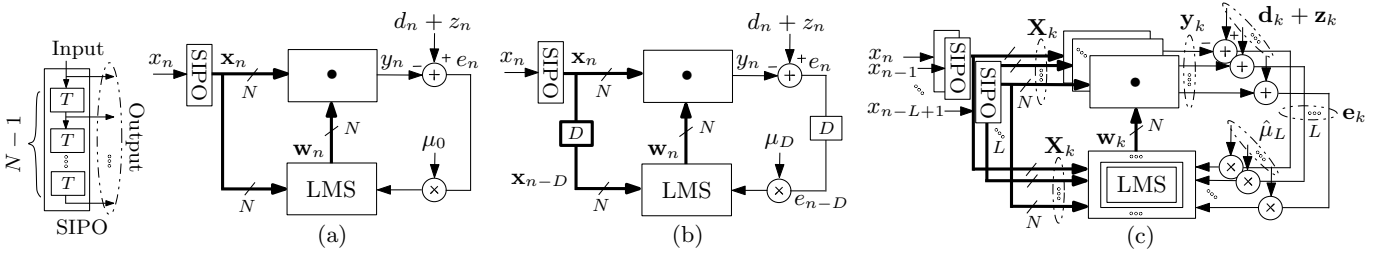}
	\caption{Top-level diagram of (a) LMS algorithm with ${\mu}_0$ as the step-size. (b) DLMS algorithm with $D$ as the number of delays and ${\mu}_D$ as the step-size. (c) BLMS algorithm with $L$ as the block size and $\hat{\mu}_L={\mu}_L/L$ as the effective step-size. $\bullet$ is the inner product operator, SIPO stands for serial-in parallel-out, $T$ indicates a sample delay, thick and thin lines represent vector and scalar quantities, respectively.}
	\label{fig11}
\end{figure*} 

\section{LMS and Related Algorithms}\label{sec:lms}
The adaptive filtering discussed here is of LMS type as presented by Widrow \textit{et al.} \cite{widrow1977stationary}. For comparison purposes of different related algorithms, we consider the system identification problem to model the unknown plant with coefficients ${\mathbf{w}}^o$. To begin with, we first explore the LMS algorithm and then its existing high-speed variants. 
\subsection{The LMS Algorithm}
At time instant $n$, the actual output $y_n$ of an $N$-tap LMS ADF can be expressed as inner product form: 
\begin{equation}\label{eq:lms output}
{{y}}_n={\mathbf{w}}^{T}_n\boldsymbol{\cdot}{\mathbf{x}}_n=\sum_{i=0}^{N-1}w_{n}(i)x_{n-i},
\end{equation}
where
\begin{equation*}
{\mathbf{x}}_n=\left[x_n, x_{n-1}, \dots x_{n-N+1}\right]^{T}
\end{equation*}
is the input vector and 
\begin{equation*}
{\mathbf{w}}_n=\left[w_n(0), w_n(1), \dots, w_n(N-1)\right]^{T}
\end{equation*}
is the filter coefficient vector. At time instant $n$, the LMS algorithm as shown in Fig.~\ref{fig11}(a) adjusts the coefficients, according to
\begin{equation}\label{eq:lms}
{\mathbf{w}}_{n+1}={\mathbf{w}}_{n}+\mu_0 e_n {\mathbf{x}}_n, 
\end{equation}
where $\mu_0$ is the step-size of the LMS algorithm, and $e_n$ is the error, which is calculated as the difference between unknown plant output $d_n$ with an i.i.d., additive noise $z_n$ and the actual output $y_n$, i.e., $e_n=d_n+z_n-y_n$. The noise $z_n$ is zero-mean, with variance $\sigma^2_z$. It is independent of any other signal in the system. 

For guaranteed convergence and algorithm stability, it is important to select an appropriate step size $\mu_0$. As per \cite{slock1993convergence}, it can be selected within the bound $\mu_{0,\mathrm{crit}}$ as
\begin{equation}\label{eq:ss_inequality_lms}
0<\mu_0< \mu_{0,\mathrm{crit}} \  \text{s.t.} \ \mu_{0,\mathrm{crit}}=\frac{2}{(N+2)\sigma^2},    
\end{equation}
where $\sigma^2$ is the input signal power. This bound can be obtained when the assumptions of {\textit{independence theory}} \cite{widrow1977stationary, mazo1979independence, feuer1985convergence, nascimento2000learning} are used. In this framework, we assume ${\mathbf{x}}_{n}$ is independent, zero mean, and jointly Gaussian. Thus, applying expectation on both sides of (\ref{eq:lms}), we have 
\begin{equation}\label{eq:lms_stoch}
\E[{\mathbf{w}}_{n+1}]=\left(I-\mu_0{\mathbf{R}}_0\right)\E[{\mathbf{w}}_{n}]+\mu_0{\mathbf{p}}_0,
\end{equation}
where ${\mathbf{{R}}}_0=\E\left[{\mathbf{x}}_{n}{\mathbf{x}}^{T}_{n}\right]$ and 
${\mathbf{{p}}}_0=\E\left[{\mathbf{x}}_{n}{{d}}_{n}\right]$. The convergence of the mean is guaranteed if $0 < \mu_0 < 2/\mathrm{tr}\left[{\mathbf{{R}}}_0\right]$, where $\mathrm{tr}[\cdot]$ is the trace operator. Using the MSE of $e_n$, i.e., $\xi_L=\E\left[e^2_n\right]$ as a performance measure leads to the best results. It follows then that the optimal set of filter coefficients  ${\mathbf{w}}^{*}={\mathbf{{R}}}_0^{-1}{\mathbf{{p}}}_0$ which is same as the Wiener filter. 

%For high-sampling rates in LMS ADF, either pipelining using DLMS algorithm \cite{long1989lms,long1992corrections} or block processing using BLMS algorithm \cite{burrus1971block, mitra1978block, clark1981block} is employed. 
\subsection{The DLMS Algorithm}
Unlike the LMS algorithm, the inputs ${\mathbf{x}}_{n}$ and the error $e_n$ in the DLMS algorithm become available after $D$ delays, as shown in Fig.~\ref{fig11}(b). It follows that the delayed error $e_{n-D}$ and inputs ${\mathbf{x}}_{n-D}$ are used to update the coefficients. The DLMS algorithm adjusts the coefficients as
\begin{equation}\label{eq:dlms}
{\mathbf{w}}_{n+1}={\mathbf{w}}_{n}+{\mu}_D e_{n-D} {\mathbf{x}}_{n-D},     
\end{equation}
where ${\mu}_D$ is the step-size of DLMS algorithm, and $e_{n-D}$ is the error obtained by the difference of unknown plant delayed output $d_{n-D}$ (with delayed noise $z_{n-D}$) and the actual delayed output $y_{n-D}$ i.e., $e_{n-D}=d_{n-D}+z_{n-D}-y_{n-D}$. The choice of step-size ${\mu}_D$ is critical as it determines the filter convergence and stability. The step-size ${\mu}_D$ can be selected within bound under first order approximation \cite{long1992corrections, solo1994adaptive} as 
 \begin{equation}\label{eq:ss_inequality_dlms}
0<\mu_D< \mu_{D,\mathrm{crit}} \  \text{s.t.} \  \mu_{D,\mathrm{crit}}=\frac{2}{(N+2+2D)\sigma^2}.     
\end{equation}

Similar to LMS, the independence assumptions also hold for DLMS. Thus, applying expectation on both sides of (\ref{eq:dlms}), we have 
\begin{equation}\label{eq:dlms_stoch}
\E[{\mathbf{w}}_{n+1}]=\E[{\mathbf{w}}_{n}]-{\mu}_D{\mathbf{{R}}}_D\E[{\mathbf{w}}_{n-D}]+{\mu}_D{\mathbf{{p}}}_D,
\end{equation}
where ${\mathbf{{R}}}_D=\E[{\mathbf{x}}_{n-D}{\mathbf{x}}^{T}_{n-D}]$ and 
${\mathbf{p}}_D=\E[{\mathbf{x}}_{n-D}{{d}}_{n-D}]$. Unlike LMS, the convergence of the mean of DLMS also depends on the covariance of filter coefficients separated by $D$ delays. By using delayed MSE of $e_{n-D}$, i.e., $\xi_D=\E\left[e^2_{n-D}\right]$, it follows that the optimal set of filter coefficients for the DLMS algorithm is given as ${\mathbf{w}}^{*}_D={\mathbf{{R}}}^{-1}_D{\mathbf{{p}}}_D$.
\subsection{The BLMS Algorithm}
Unlike the LMS and DLMS algorithms, BLMS processes $L$ inputs and produces $L$ outputs in one iteration by fixing the coefficients for $L$ samples in block iteration $k$, i.e., ${\mathbf{w}}_k={\mathbf{w}}_n$ such that $k=nL$. A typical configuration of the BLMS algorithm is shown in Fig.~\ref{fig11}(c). At block iteration, $k$, (\ref{eq:lms output}) can be expressed as a block of inner products:
\begin{equation}\label{eq:blms_output}
{\mathbf{y}}_{k}={\mathbf{X}}_k\boldsymbol{\cdot}{\mathbf{w}}_k,
\end{equation}
where 
\begin{equation*}
{\mathbf{y}}_k=\left[{{y}}_{(k-1)L+1}, {{y}}_{(k-1)L+2},\dots, {{y}}_{kL}\right]^{T}
\end{equation*}and 
\begin{equation*}
{\mathbf{X}}_k=\left[{\mathbf{x}}_{(k-1)L+1}, {\mathbf{x}}_{(k-1)L+2},\dots,{\mathbf{x}}_{kL}\right]^{T}.
\end{equation*}

The error block ${\mathbf{e}}_k$ is computed as the difference between unknown plant block output 
\begin{equation*}
{\mathbf{d}}_{k}=\left[d_{(k-1)L+1}, d_{(k-1)L+2}, \dots, d_{kL}\right]^{T}
\end{equation*}
with noise
\begin{equation*}
{\mathbf{z}}_{k}=\left[z_{(k-1)L+1}, z_{(k-1)L+2}, \dots, z_{kL}\right]^{T}
\end{equation*}
and actual block output ${\mathbf{y}}_k$ as 
\begin{equation}
{\mathbf{e}}_k={\mathbf{d}}_k+{\mathbf{z}}_k-{\mathbf{y}}_k=\left[e_{(k-1)L+1}, e_{(k-1)L+2},\dots,e_{kL}\right]^{T}.
\end{equation} Using ${\mathbf{e}}_k$, the coefficients ${\mathbf{w}}_{k}$ are updated as 
\begin{equation}\label{eq:blms}
{\mathbf{w}}_{k+1}={\mathbf{w}}_{k}+\frac{{\mu}_L}{L}{\mathbf{X}}^{T}_{k}{\mathbf{e}}_{k},
\end{equation}
where ${\mu}_L$ is the step-size of the BLMS algorithm whose choice is critical to determine its convergence and stability \cite{clark1981block}. In \cite{feuer1985performance, lee2012step}, a bound for the step-size ${\mu}_L$ is given as 
\begin{equation}\label{eq:ss_inequality_Llms}
0<\mu_L<  \mu_{L,\mathrm{crit}} \  \text{s.t.} \  \mu_{L,\mathrm{crit}}=\frac{2L}{(N+L+1)\sigma^2}.      
\end{equation}
\begin{figure}
	\centering
	\includegraphics[width=0.90\linewidth]{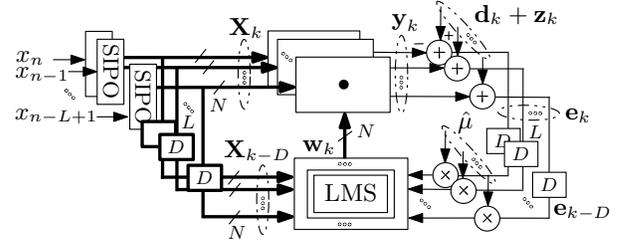}
	\caption{Top-level diagram of DBLMS algorithm with $\hat{\mu}={\mu}/L$ is the effective step-size.}
	\label{fig2n}
\end{figure}%

Again, the previous independence assumptions hold for the BLMS algorithm. Thus, applying expectation on both sides of (\ref{eq:blms}), we get
\begin{equation}\label{eq:blms_stoch}
\E[{\mathbf{w}}_{k+1}]=\left(I-{{{\mu}_L}}{\mathbf{{R}}}_L\right)\E[{\mathbf{w}}_{k}]+{{{\mu}_L}}{{\mathbf{{p}}}_L},   
\end{equation}
where ${\mathbf{{R}}}_L=\E\left[{\mathbf{X}}^{T}_{k}{\mathbf{X}}_{k}\right]$ and ${\mathbf{{p}}}_L=\E\left[{\mathbf{X}}^T_{k}{\mathbf{d}}_{k}\right]$. Likewise, the convergence of the mean is guaranteed if $0 < {\mu}_L <  1/\mathrm{tr}\left[{\mathbf{{R}}}_L\right]$. By
using block MSE of ${\mathbf{e}}_k$, i.e., $\xi_{B}=\frac{1}{L}\E[{\mathbf{e}}^T_k{\mathbf{e}}_k]$, it follows then that the optimal set of filter coefficients for BLMS algorithm can be given as ${\mathbf{w}}^{*}_L={{{\mathbf{{R}}}^{-1}_L}}{\mathbf{{p}}}_L$. For stationary inputs, ${\mathbf{{R}}}_L$ and ${\mathbf{{p}}}_L$ are simply $L$ times of ${\mathbf{R}}_0$ and ${\mathbf{p}}_0$ respectively. Thus, they must have similar properties as those of ${\mathbf{R}}_0$ and ${\mathbf{p}}_0$ which are listed below:
\begin{enumerate}
	\item ${\mathbf{{R}}}_L$ and ${\mathbf{{R}}}_0$ are symmetric (Hermitian) and positive definite,
\item ${\mathbf{{R}}}_L$ and ${\mathbf{{R}}}_0$ posses $N$ linearly independent eigenvectors and can be reduced to a diagonal form by a similarity transformation \cite{fu2006simultaneous},
\item Eigenvalues of ${\mathbf{{R}}}_L$ are real, positive and $\times L$ of ${\mathbf{{R}}}_0$. 
\end{enumerate}

\section{The DBLMS Algorithm}
%\subsection{Block Diagram}
The top-level diagram for the DBLMS algorithm is shown in Fig.~\ref{fig2n}. The symbols used in the analysis of DBLMS are the same as the BLMS algorithm except that subscript $L$ is dropped for the sake of brevity and simplicity. For clarity, we list the symbols used for the analysis of the DBLMS algorithm in Table~\ref{tab1}. 
\begin{table}
	\caption{Symbols Used in the Analysis of the DBLMS Algorithm.}
	\centering
	\begin{tabular}{|l|c|}
		\hline
		\textbf{Quantity}  & \textbf{Symbol} \\ \hline
		Filter taps & $N$ \\
		Block size & $L$ \\
		Adaptation loop delays & $D$ \\
		Speedup          & $S=(D+1)L$          \\ 
		Input matrix         & ${\mathbf{X}}_{k}$          \\ 
		Error vector        & ${\mathbf{e}}_{k}$          \\ 
		Step-size         & $\mu$          \\ 
		Effective step-size         & ${\hat\mu}$          \\ 		
		Effective critical step-size (or bound)         & $\hat{\mu}_{\mathrm{crit}}$          \\ 
		Effective optimum step-size         & $\hat{\mu}_{\mathrm{opt}}$          \\ 
		Scaled step-size         & $\hat{\hat\mu}$          \\ 
		Delayed block mean squared error         & $\xi$          \\ 
		Delayed block excess MSE         & $\xi_{\mathrm{ex}}$          \\ 
		Delayed block minimum MSE         & $\xi_{\mathrm{min}}$          \\ 
		Gradient         & $\nabla_{k}$          \\ 
		Unbiased gradient         & $\tilde\nabla_{k}$          \\ 
		Eigenvalues         & $\lambda_{i}$ \\ % (i=0,1,\dots,N-1)$          \\ 
		Diagonal vector         & ${\mathbf{\Lambda}}$          \\ 
		Root mean square of $\lambda_{i}$        & $\lambda_{\mathrm{rms}}$          \\ 
		Average of $\lambda_{i}$          & %$\lambda_{\mathrm{avg}}$ or $\sigma^2$         \\ 
		$\sigma^2$         \\ 
		Misadjustment         & $M=\xi_{\mathrm{ex}}/\xi_{\mathrm{min}}$          \\ 
		Time constant         & $\tau$          \\ 
		Mapping matrix & ${\mathbf{S}}_{\hat{\hat\mu} L,D}$ \\
		Slope factor         & $\alpha_{\hat{\hat{\mu}}L,D}$          \\  
	%	???          & $\rho = \lambda^2_{\mathrm{rms}}/\lambda^2_{\mathrm{avg}}$          \\ 
		Kurtosis          & $\nu$          \\ \hline
	%	???          & $P = \rho(N+(\nu-1)L)$          \\ 
	%	???          & $K = \rho(N(1+1/M)+2L)$          \\ \hline 
		
	\end{tabular}\label{tab1}
\end{table}

\subsection{Optimum Filter Coefficients}
The original Wiener filter can also be extended to the delayed block input case, as shown in Fig.~\ref{fig2n}. Notably, the coefficients in the DBLMS algorithm are updated with a delayed block of inputs and errors. To determine the optimum filter coefficients, we define the error ${\mathbf{e}}_{k-D}$ as
\begin{IEEEeqnarray}{rCl}
{\mathbf{e}}_{k-D} & = & {\mathbf{d}}_{k-D}+{\mathbf{z}}_{k-D}-{\mathbf{y}}_{k-D} \label{eq:dblms_error} \\
& = & \left[e_{(k-D-1)L+1}, e_{(k-D-1)L+2}, \dots, e_{(k-D)L}\right]^T. \nonumber
\end{IEEEeqnarray}
The desired delayed block input
\begin{equation*}
{\mathbf{d}}_{k-D}=\left[d_{(k-D-1)L+1}, d_{(k-D-1)L+2}, \dots, d_{(k-D)L}\right]^T
\end{equation*}
is obtained by delaying the desired block input ${\mathbf{d}}_{k}$ through $D$ delays. Similarly, the noise ${\mathbf{z}}_{k-D}$ can be obtained by delaying the desired block noise ${\mathbf{z}}_{k}$ through $D$ delays. The delayed output block from (\ref{eq:blms_output}) can be written as  
\begin{equation}\label{eq:dblms_output}
{\mathbf{y}}_{k-D}={\mathbf{X}}_{k-D}\boldsymbol{\cdot}{\mathbf{w}}_{k-D}.
\end{equation}

For clarity, we show the computation of the inner product (\ref{eq:dblms_output}) by dropping the time index from ${\mathbf{w}}_{k-D}$ for $N=3$, $D=1$, and $L=2$:
\begin{equation}\label{eq:dblms_demonstration}
\begin{tabular}{l}
${\mathbf{X}}_{-1}\left\{\lefteqn{\phantom{\begin{matrix} x_{-1}\\ x_{0}\ \end{matrix}}}\right.$\\
\hspace{+0.7em}${\mathbf{X}}_{0}\left\{\lefteqn{\phantom{\begin{matrix} x_1\\ x_2\ \end{matrix}}} \right.$\\
\hspace{+0.7em}${\mathbf{X}}_{1}\left\{\lefteqn{\phantom{\begin{matrix} x_3\\ x_4\ \end{matrix}}} \right.$\\
$\lefteqn{\phantom{\begin{matrix} 0 \end{matrix}}} $
\end{tabular}
\left[\phantom{\begin{matrix}x-1\\ x_0\\ x_1\\ x_2\\ x_3\\ x_4\\ \cdot\cdot\\ \end{matrix}}
\hspace{+0.5em}
\right.
\hspace{-2.5em}
\begin{matrix}
x_{-1} & 0 & 0  \\
x_{0} & x_{-1} & 0  \\ \hdashline
x_1 & x_0 & x_{-1}  \\
x_2 & x_1 & x_0  \\\hdashline
x_3 & x_2 & x_1  \\
x_4 & x_3 & x_2  \\\hdashline
\cdot\cdot & \cdot\cdot & \cdot\cdot  
\end{matrix}
\hspace{-2.0em}
\left.\phantom{\begin{matrix}0\\0\\x-1\\ x_0\\ x_1\\ x_2\\  \cdot\cdot\\ \end{matrix}}\right]
\boldsymbol{\cdot} \underbrace{\left[\begin{matrix}
	w_1  \\
	w_2 \\
	w_3 \\
	\end{matrix}\right]}_{{\mathbf{W}}}=
\left[\begin{matrix}
y_{-1}  \\
y_0 \\ \hdashline
y_1 \\
y_2 \\ \hdashline
y_3 \\
y_4 \\\hdashline
\cdot\cdot 
\end{matrix}\right]
\begin{tabular}{l}
$\left.\lefteqn{\phantom{\begin{matrix} y_{-1}\\ y_0\ \end{matrix}}}\right\}{\mathbf{y}}_{-1}$\\
$\left.\lefteqn{\phantom{\begin{matrix} y_1\\ y_2\ \end{matrix}}} \right\}{\mathbf{y}}_{0}$\\
$\left.\lefteqn{\phantom{\begin{matrix} y_3\\ y_4\ \end{matrix}}} \right\}{\mathbf{y}}_{1}$\\
$\lefteqn{\phantom{\begin{matrix} 0 \end{matrix}}} $
\end{tabular}.
\end{equation}

The block of inputs and outputs is delayed by the unit amount compared to the zero delayed block method (\ref{eq:blms_output}). Using (\ref{eq:dblms_error}), we estimate the delayed block MSE (DBMSE) $\xi$ as a cost function that the DBLMS algorithm minimizes as 
\begin{equation}\label{eq:dblms_mse}
\xi=\frac{1}{L}\E\left[{\mathbf{e}}^T_{k-D}{\mathbf{e}}_{k-D}\right]=\E\left[\frac{1}{L}\sum_ {n=(k-D-1)L+1}^{(k-D)L}e^2_{n}\right].   
\end{equation}
From (\ref{eq:dblms_mse}), it is clear that the DBMSE is the expected value of an estimate of squared error over one delayed block. Further, it combines the information corresponding to delays and blocks into a single value in the least squares sense \cite{gersho1969adaptive}. Using (\ref{eq:dblms_error}), we can further express (\ref{eq:dblms_mse}) as
\begin{IEEEeqnarray}{-rCl}\label{eq:dblms_mse_other}
	L\xi&=&\sigma^2_z+\E\left[{\mathbf{d}}^T_{k-D}{\mathbf{d}}_{k-D}\right]-\E\left[{\mathbf{d}}^T_{k-D}{\mathbf{X}}_{k-D}\right]{\mathbf{w}}\nonumber\\ 
	&&-{\mathbf{w}}^T\E\left[{\mathbf{d}}^T_{k-D}{\mathbf{X}}_{k-D}\right]+{\mathbf{w}}^{T}\E\left[{\mathbf{X}}^T_{k-D}{\mathbf{X}}_{k-D}\right]{\mathbf{w}}.
\end{IEEEeqnarray}

The terms containing ${\mathbf{z}}_{k-D}$ were suppressed as their expected values are equal to zero in (\ref{eq:dblms_mse_other}), except with a second-order term of 
${\mathbf{z}}_{k-D}$ which corresponds to variance $\sigma^2_z=\E\left[{\mathbf{z}}^T_{k-D}{\mathbf{z}}_{k-D}\right]$. We define the correlation matrices for the DBLMS algorithm as
\begin{equation}
{\mathbf{R}}=\E\left[{\mathbf{X}}^{T}_{k-D}{\mathbf{X}}_{k-D}\right] \text{ and }{\mathbf{{p}}}=\E\left[{\mathbf{X}}^T_{k-D}{\mathbf{d}}_{k-D}\right].
\end{equation}
Using these definitions and invoking the stationarity, we can re-write (\ref{eq:dblms_mse_other}) as
%where ${\mathbf{d}}'_{k-D}={\mathbf{d}}_{k-D}+{\mathbf{z}}_{k-D}$. 
\begin{equation}\label{eq:dblms_mse_simplified}
\xi=\frac{1}{L}\left(\E\left[{\mathbf{d}}^T_{k-D}{\mathbf{d}}_{k-D}\right]-2{\mathbf{p}}^T{\mathbf{w}}+{\mathbf{w}}^T{\mathbf{R}}{\mathbf{w}}+\sigma^2_z\right).
\end{equation}

Defining 
\begin{equation}\label{eq:desired_variance}
\sigma^2_d=\E\left[{\mathbf{d}}^T_{k-D}{\mathbf{d}}_{k-D}\right]
\end{equation}
where $\sigma^2_d$ is the variance of the delayed block desired signal. It is clear from (\ref{eq:dblms_mse_simplified}) that the DBMSE is the same as the delayed MSE of DLMS for the stationary inputs, i.e., $\xi=\xi_D$, it follows then that their optimal set of filter coefficients are also equal, i.e., ${\mathbf{w}}^{*}={\mathbf{w}}^{*}_D$. One can find the expression of minimum DBMSE by extending the principle of orthogonality which says that coefficient vector ${\mathbf{w}}^{*}$ minimizes the DBMSE for which ${\mathbf{e}}_{k-D}$ is orthogonal to ${\mathbf{X}}_{k-D}$. Hence, the minimum DBMSE expression can be given as
\begin{equation}\label{eq:min DBMSE}
\xi_{\mathrm{min}}=\frac{1}{L}\E\left[{\mathbf{d}}^T_{k-D}{\mathbf{e}}_{k-D}\right]=\frac{1}{L}\left(\sigma^2_d-{\mathbf{p}}^T{\mathbf{w}}^{*}\right).    
\end{equation}
In order to calculate $\xi_{\mathrm{min}}$, it is required to obtain ${\mathbf{w}}^{*}$. This will be discussed later in a subsection.
\subsection{Coefficient Update Equation}
Analogous to LMS, DLMS, and BLMS adaptive filtering, a delayed block algorithm can be obtained by solving for the Wiener coefficient vector in real time using gradient search method \cite{widrow1971adaptive,widrow1976comparison}. As stated, in the DLMS algorithm, it is desired that the coefficients are updated by the delayed inputs and errors. In contrast, the BLMS algorithm keeps the coefficients fixed in every new block of data. Thus, DBLMS in gradient form can be expressed as 
\begin{equation}\label{eq:dblms}
{\mathbf{w}}_{k+1}={\mathbf{w}}_{k}-\frac{1}{2}\mu \nabla_{k}, 
\end{equation}
where $\mu$ is the step size of DBLMS algorithm and $\nabla_{k}$ is the gradient of $\xi$ at block iteration $k$. Formally, the gradient is expressed with respect to the coefficients:
\begin{equation}\label{eq:dblms_gradient}
\nabla_{k}\triangleq\frac{\partial \xi}{\partial {\mathbf{w}}}=\left.\frac{1}{L}\frac{\partial\E[{\mathbf{e}}^T_{k-D}{\mathbf{e}}_{k-D}]}{\partial{\mathbf{w}}}\right|_{{\mathbf{w}}={\mathbf{w}}_k}.  
\end{equation}

An estimate, $\tilde{\nabla}_{k}$,  of the gradient $\nabla_{k}$ is
\begin{equation}\label{eq:dblms_grad_unbiased}
\tilde{\nabla}_{k}=\frac{1}{L}\frac{\partial{\left[{\mathbf{e}}^T_{k-D}{\mathbf{e}}_{k-D}\right]}}{\partial{\mathbf{w}}_k}=-\frac{2}{L}{\mathbf{X}}^{T}_{k-D}{\mathbf{e}}_{k-D}. 
\end{equation}
Using this unbiased gradient estimate in the coefficient adjust algorithm (\ref{eq:dblms}), gives the DBLMS algorithm as
\begin{equation}\label{eq:dblms_final}
{\mathbf{w}}_{k+1}={\mathbf{w}}_{k}+\frac{\mu}{L}{\mathbf{X}}^{T}_{k-D}{\mathbf{e}}_{k-D}.
\end{equation}

Note that the coefficient update term in (\ref{eq:dblms_final}) is an average of $L$ DLMS-like terms $e_{n-D}{\mathbf{X}}_{n-D}$ generated by a block of data. The choice of delay and block size is important, and (\ref{eq:dblms_final}) reveals that the algorithm is valid for any $D$ and $L$. However, algorithm stability and convergence must be examined. 

\subsection{Development of DBMSE in Diagonal Form}
To examine algorithm stability and convergence, it is required to develop the MSE expression of the DBLMS algorithm in the diagonal form \cite{fu2006simultaneous}. Recall (\ref{eq:dblms_mse_simplified}) and (\ref{eq:desired_variance}), we simplify the DBMSE as 
\begin{equation}\label{eq:dblms_mse_simplify}
\xi=\frac{1}{L}(\sigma^2_d+\sigma^2_z-2{\mathbf{{p}}}^T{\mathbf{w}}+{\mathbf{w}}^T{\mathbf{{R}}}{\mathbf{w}}).
\end{equation}
$\tilde{\nabla}$ in diagonal form using (\ref{eq:dblms_grad_unbiased}) with respect to ${\mathbf{w}}$ is
\begin{equation}\label{eq:dblms_gradient_simp}
\tilde{\nabla}=\frac{1}{L}(-2{\mathbf{{p}}}+2{\mathbf{{R}}}{\mathbf{w}}).    
\end{equation}
Using (\ref{eq:dblms_mse_simplified}), one can find an alternative and useful expression to calculate the DBMSE as
\begin{IEEEeqnarray}{rCl}\label{eq:dblms_mse_simp1}
	\xi&=&\xi_{\mathrm{min}}+\frac{1}{L}\E\left[({\mathbf{w}}-{\mathbf{w}}^{*})^T{\mathbf{{R}}}({\mathbf{w}}-{\mathbf{w}}^{*})\right]\nonumber \\
	&=&\xi_{\mathrm{min}}+\frac{1}{L}\E\left[{\mathbf{v}}^T{\mathbf{{R}}}{\mathbf{v}}\right];\hspace{0.2cm}{\mathbf{v}}\triangleq ({\mathbf{w}}-{\mathbf{w}}^{*}), 
\end{IEEEeqnarray}
where ${\mathbf{v}}$ is the coefficient error vector at block iteration $k$. 

Since ${\mathbf{{R}}}$ is symmetric and positive definite, it can be diagonalized using a similarity transformation \cite{fu2006simultaneous}, ${\mathbf{{R}}}={\mathbf{Q}}{\mathbf{\Lambda}}{\mathbf{Q}}^{-1}={\mathbf{Q}}{\mathbf{\Lambda}}{\mathbf{Q}}^{T}$, where ${\mathbf{Q}}$ is the orthonormal $({\mathbf{Q}}^{T}={\mathbf{Q}}^{-1})$ modal matrix of ${\mathbf{{R}}}$, and ${\mathbf{\Lambda}}$ is the diagonal matrix of eigenvalues $\lambda_{i}$ ($i=0,1,\dots,N-1$)  of ${\mathbf{{R}}}$. Thus, (\ref{eq:dblms_mse_simp1}) can be re-written as
\begin{equation}\label{eq:dblms_mse_transformed}
\xi=\xi_{\mathrm{min}}+\frac{1}{L}\E\left[{{\mathbf{v}}^{T}{\mathbf{Q}}{\mathbf{\Lambda}}{\mathbf{Q}}^{-1}{\mathbf{v}}}\right].   
\end{equation}
Using linear transformation, a new variable $\tilde{\mathbf{v}}$ is introduced as
\begin{equation}\label{eq:dblms_trans_variables}
\tilde{\mathbf{v}}\triangleq {\mathbf{Q}}^{-1}{\mathbf{v}}={\mathbf{Q}}^{T}{\mathbf{v}}\hspace{0.1cm} \mathrm{and} \hspace{0.1cm} {\mathbf{v}}={\mathbf{Q}}\tilde{\mathbf{v}}.  
\end{equation}

Similarly, the delayed block input matrix is transformed to $\tilde{{\mathbf{X}}}_{k-D}={{\mathbf{Q}}}({{\mathbf{X}}}_{k-D})$. Based on this new coordinate system, (\ref{eq:dblms_mse_transformed}) may be precisely written as
\begin{equation}\label{eq:dblms_mse_simp_final}
\xi=\xi_{\mathrm{min}}+\frac{1}{L}\E\left[{{\tilde{\mathbf{v}}}^{T}\mathbf{\Lambda}\tilde{\mathbf{v}}}\right].    
\end{equation}
According to (\ref{eq:dblms_mse_simp_final}), DBMSE at any iteration $k$ depends on the composite product  $\E\left[{{\tilde{\mathbf{v}}}^{T}\mathbf{\Lambda}\tilde{\mathbf{v}}}\right]/L$ over one delayed block.

\subsection{Mean Coefficient Behaviour and Coefficient-Error Vector}
For guaranteed convergence and algorithm stability, it is important to analyze and select an appropriate $\mu$ for a given choice of $L$ and $D$. Similar to LMS, DLMS, and BLMS, independence assumptions hold for DBLMS i.e., the elements of input matrix ${\mathbf{X}}_{k-D}$ are uncorrelated in time, zero mean, and jointly Gaussian. 
Thus, the mean coefficient behaviour of the DBLMS ADF is determined by applying the expectation on both sides of (\ref{eq:dblms_final}) as
\begin{equation}\label{eq:dblms_stoch}
\E[{\mathbf{w}}_{k+1}]=\E[{\mathbf{w}}_{k}]-\frac{\mu}{L}{\mathbf{{R}}}\E[{\mathbf{w}}_{k-D}]+\frac{\mu}{L}{\mathbf{{p}}}.
\end{equation}

Like DLMS, the convergence of the mean of DBLMS also depends on the covariance of the coefficients separated by past $D$ delays.
From (\ref{eq:dblms_stoch}), we can find the steady-state coefficient vector, if the DBLMS algorithm convergence is assumed. Then, one can obtain the steady-state filter coefficients from the condition $\displaystyle \lim_{k \to \infty} \E[{\mathbf{w}}_{k}]=\E[{\mathbf{w}}_{k-D}]={\mathbf{w}}^{*}$ as 
\begin{equation}\label{eq:opt_coeff}
{\mathbf{w}}^{*}={\mathbf{{R}}}^{-1}{\mathbf{{p}}}.    
\end{equation}

As expected, ${\mathbf{w}}^{*}$ converges to the mean of DBMSE in (\ref{eq:dblms_mse_simplified}), if $\xi=\xi_{\mathrm{min}}$. These estimates are usually not sufficient as it only involves first-order moments. Further, no guarantee that the mean of coefficients will converge within finite variance, as indicated by the second term of (\ref{eq:dblms_mse_simp_final}). Therefore, it is necessary to carry out the analysis with second-order statistics. As per (\ref{eq:dblms_mse_simp1}), it is useful to analyze the coefficients vector in terms of their error vector as it could provide better estimates. Thus, the coefficient update equation in terms of $\tilde{{\mathbf{v}}}_{k}$ can be given as 
\begin{IEEEeqnarray}{rCl}\label{eq:dblms_transf_other}
	\tilde{\mathbf{v}}_{k+1}&=&\tilde{\mathbf{v}}_{k}-\frac{\mu}{L}\tilde{\mathbf{X}}^T_{k-D}\tilde{\mathbf{X}}_{k-D}\tilde{\mathbf{v}}_{k-D}+\frac{\mu}{L}
	\tilde{\mathbf{X}}^T_{k-D}{\mathbf{z}}_{k-D}\nonumber \\
	&&-\frac{\mu}{L}
	\tilde{\mathbf{X}}^T_{k-D}\tilde{\mathbf{X}}_{k-D}\tilde{\mathbf{w}}^{*}+\frac{\mu}{L}
		\tilde{\mathbf{X}}^T_{k-D}{\mathbf{d}}_{k-D}.
	\end{IEEEeqnarray}

Defining effective step-size $\hat\mu={\mu}/{L}$ and ${\mathbf{e}}_{\mathrm{o},k-D}={\mathbf{d}}_{k-D}-\tilde{\mathbf{X}}_{k-D}\tilde{\mathbf{w}}^{*}$ in (\ref{eq:dblms_transf_other}), we can simplify it as
		\begin{equation}\label{eq:dblms_transf_simp}
		\tilde{\mathbf{v}}_{k+1}=\tilde{\mathbf{v}}_{k}-\hat\mu\tilde{\mathbf{X}}^{T}_{k-D}\tilde{\mathbf{X}}_{k-D}\tilde{\mathbf{v}}_{k-D}+\hat\mu \tilde{\mathbf{X}}^T_{k-D}({\mathbf{e}}_{\mathrm{o},k-D}+{\mathbf{z}}_{k-D}).
	\end{equation}
	The update of coefficient error vector depends on $\tilde{\mathbf{v}}_{k}$, $\tilde{\mathbf{v}}_{k-D}$,  ${\mathbf{e}}_{\mathrm{o},k-D}$, ${\mathbf{z}}_{k-D}$ and ${\mathbf{X}}_{k}$. 

	\section{Convergence of the DBLMS Algorithm} 
	In this Section, the mathematical expressions for the convergence of the DBLMS algorithm in the steady and transient states are derived. 
	\subsection{Steady-State Analysis}
 Defining the second term of (\ref{eq:dblms_mse_simp_final}) as the excess DBMSE, i.e., $\xi_{k,{\mathrm{ex}}}$ at iteration $k$, we obtain
	\begin{equation}\label{eq: excess_MSE_def}
	L\xi_{k,\mathrm{ex}}=L(\xi_{k}-\xi_{\mathrm{min}})=\E\left[{{\tilde{\mathbf{v}}_k}^{T}\mathbf{\Lambda}\tilde{\mathbf{v}}_k}\right]
	\end{equation}
    The definition of (\ref{eq: excess_MSE_def}) can be generalized with respect to past coefficient error vectors ${\tilde{\mathbf{v}}_{k-r}}$ and ${\tilde{\mathbf{v}}_{k-s}}$ as 
	\begin{equation}\label{eq:exc_MSE_gen}
	L\xi_{k,rs}=
	\E\left[{{\tilde{\mathbf{v}}_{k-r}}^{T}{\mathbf{\Lambda}}\tilde{\mathbf{v}}_{k-s}}\right].
	\end{equation}
	Thus, one can deduce $\xi_{k,\mathrm{ex}}$ as
	\begin{equation}\label{eq: relation}
	\xi_{k,\mathrm{ex}}=\xi_{k,00}.    
	\end{equation}

	The following properties are used in the DBMSE analysis:
	\begin{IEEEeqnarray}{rll}
		\IEEEyesnumber\label{eq:both3} \IEEEyessubnumber*
		\xi_{k-D-1,rs} = & \hspace{0.1cm} \xi_{k-1,(r-D)(s-D)}, \label{eq:sub1}\\
		\xi_{k-1,0D} = & \hspace{0.1cm} \xi_{k-1,D0},  \label{eq:sub2}\\
			\xi_{k-1,(r+1)s} = & \hspace{0.1cm} \xi_{k-s-1,0(r-s+1)}.  \label{eq:sub3}
	\end{IEEEeqnarray}

	By expanding the product terms in (\ref{eq: excess_MSE_def}) using (\ref{eq:dblms_transf_simp}), taking the expectation on the result, utilizing the relations in (\ref{eq: relation}) and (\ref{eq:sub2}) with some algebraic manipulation (see Appendix \ref{Appendix:A}), and defining $k_D=k-D-1$ for brevity, we have  
	\begin{IEEEeqnarray}{rCl}
		%\IEEEeqnarraymulticol{3}{L}{L\xi_{k,00}} \\ \nonumber 
		L\xi_{k,00}&=&L\xi_{k-1,00} -2\hat{\mu}\E\left[\tilde{{\mathbf{v}}}^T_{k_D}{\mathbf{\Lambda}}^2\tilde{{\mathbf{v}}}_{k_D+D}\right]\nonumber\\
		&&+\hat{\mu}^2\E\left[\tilde{{\mathbf{v}}}^T_{k_D}{\tilde{{\mathbf{X}}}^T_{k_D}}{\tilde{{\mathbf{X}}}_{k_D}}{\mathbf{\Lambda}}{\tilde{{\mathbf{X}}}^T_{k_D}}{\tilde{{\mathbf{X}}}_{k_D}}\tilde{{\mathbf{v}}}_{k_D}\right]\nonumber\\
		&&+\hat{\mu}^2\E\left[{\mathbf{e}}^T_{{\mathrm{o}},k_D}{\tilde{{\mathbf{X}}}_{k_D}}{\mathbf{\Lambda}}{\tilde{{\mathbf{X}}}^T_{k_D}}{\mathbf{e}}_{{\mathrm{o}},k_D}\right]\nonumber\\
		&&+\hat{\mu}^2\E\left[{\mathbf{z}}^T_{k_D}{\tilde{{\mathbf{X}}}_{k_D}}{\mathbf{\Lambda}}{\tilde{{\mathbf{X}}}^T_{k_D}}{\mathbf{z}}_{k_D}\right], \label{eq:dblms_mse_expand}
	\end{IEEEeqnarray}

Equation (\ref{eq:dblms_mse_expand}) contains four non-trivial terms which will be analyzed individually for the sake of clarity. The middle term in (\ref{eq:dblms_mse_expand}), excluding $\tilde{{\mathbf{v}}}^T_{k_D}$ and $\tilde{{\mathbf{v}}}_{k_D}$, can be written as
\begin{IEEEeqnarray}{rCl}
	\IEEEeqnarraymulticol{3}{l}{\E\left[{\tilde{{\mathbf{X}}}^T_{k_D}}{\tilde{{\mathbf{X}}}_{k_D}}{\mathbf{\Lambda}}{\tilde{{\mathbf{X}}}^T_{k_D}}{\tilde{{\mathbf{X}}}_{k_D}}\right]} \nonumber\\ 
	& = & \E\left[{\tilde{{\mathbf{X}}}^T_{k_D}}\left(\sum\limits_{j=1}^{L}\sum\limits_{i=1}^{N}\lambda_{i}{\tilde{x}}^2_{k_DL+j+i}\right){\tilde{{\mathbf{X}}}_{k_D}}\right], \label{eq:dblms_start}
\end{IEEEeqnarray}

Define the symmetric Gram matrix ${\tilde{{\mathbf{Y}}}_{k_D}}={\tilde{{\mathbf{X}}}^T_{k_D}}{\tilde{{\mathbf{X}}}_{k_D}}$. Each element, ${\tilde{{\mathbf{Y}}}_{k_D}}(u,v)$, can be expressed as 
	\begin{equation}\label{eq:dblms_product}
{\tilde{{\mathbf{Y}}}_{k_D}}(u,v)=\begin{cases}
	\displaystyle\sum\nolimits_{j=1}^{L}{\tilde{x}}^2_{k_DL+j+u}, & \text{$u=v$}.\\
	\displaystyle\sum\nolimits_{j=1}^{L}{\tilde{x}}_{k_DL+j+u}{\tilde{x}}_{k_DL+j+v}, & \text{$u\neq v$}.
	\end{cases}
	\end{equation}
A single element of the composite matrix in (\ref{eq:dblms_start}) is written as
	\begin{equation}\label{eq:dblms_next}
	\E\left[\sum\limits_{j=1}^{L}\sum\limits_{i=1}^{N}\lambda_{i}{\tilde{x}}^2_{k_DL+j+i}{\tilde{{\mathbf{Y}}}_{k_D}}(u,v)\right].    
	\end{equation}	
%	\marginpar{Isn't there a relation? Is $\tilde{x}$ an element of the matrix? Two indices? You mean to say the relation between $\tilde{x}$ across different rows?? I have included a more detailed analysis for clarity}
	Using (\ref{eq:dblms_product}), we can re-write (\ref{eq:dblms_next}) as 
	\begin{IEEEeqnarray}{rCl}
%		\IEEEeqnarraymulticol{3}{l}{L\E\left[\delta_{uv}\sum\nolimits_{i=1}^{N}\lambda_{i}\left(\tilde{x}^2_{k_DL+p+u}\right){\tilde{x}}^2_{k_DL+i}\right]}\nonumber\\
		\IEEEeqnarraymulticol{3}{l}{\E\left[\delta_{uv}\sum\limits_{j=1}^{L}\left(\sum\limits_{i=1}^{N}\lambda_{i}{\tilde{x}}^2_{k_DL+j+i}\right)\left(\sum\limits_{j=1}^{L}{\tilde{x}}^2_{k_DL+j+u}\right)\right]}\nonumber\\
		&=&\delta_{uv}\left[\sum\limits_{i=1}^{N}\lambda_{i}\left(\E\left[\sum\limits_{j=1}^{L}{\tilde{x}}^2_{k_DL+j+i}\right]\E\left[\sum\limits_{j=1}^{L}{\tilde{x}}^2_{k_DL+j+u}\right]\right)\right.   \nonumber\\
		&&\left.+(\nu_u-1)\lambda_{u}\left(\E\left[\sum\limits_{j=1}^{L}{\tilde{x}}^2_{k_DL+j+u}\right]\right)^2\right]\nonumber\\
		&=&\delta_{uv}\left[L\left(\sum\limits_{i=1}^{N}{\lambda^2_{i}}\right)\lambda_{u}+L^2(\nu_u-1)\lambda^3_{u}\right],\label{eq:dblms_mse_expand_7th_simp1}
	\end{IEEEeqnarray}
	where $\delta_{uv}$ is the Kronecker delta function and $\nu_u$ is kurtosis defined by $\nu_u=\E\left[{\tilde{x}}^4_u\right]/\left(\E\left[{\tilde{x}}^2_u\right]\right)^2$. To simplify the analysis, all $\nu_u$ are considered to be the same, i.e., $\nu_u=\nu$. This is true if $x$ (so $\tilde{x}$) has a Gaussian distribution which results in $\nu=3$. 
	
	Re-expressing (\ref{eq:dblms_mse_expand_7th_simp1}) in the original matrix form as 
		% $L^2\nu_u=\E\left[{\tilde{x}}^4_{u}\right]/\left(\E\left[{\tilde{x}}^2_{u}\right]\right)^2$. 
	%\begin{equation}
	%	 \nu_u=\E\left[\left(\sum\nolimits_{L}{\tilde{x}}^2_{u}\right)^2\right]\Big/\left(\E\left[\left(\sum\nolimits_{L}{\tilde{x}}^2_{u}\right)\right]\right)^2.
	%\nu_u=\frac{\E\left[\left(	\displaystyle\sum\nolimits_{j=1}^{L}{\tilde{x}}^2_{k_DL+j+u}\right)^2\right]}{\left(\E\left[\left(	\displaystyle\sum\nolimits_{j=1}^{L}{\tilde{x}}^2_{k_DL+j+u}\right)\right]\right)^2}.
	%\end{equation}
	\begin{equation}\label{eq:dblms_mse_expand_7th_simp2}
	LN\lambda^2_{\mathrm{rms}}{\mathbf{\Lambda}}+L^2(\nu-1){\mathbf{\Lambda}}^3,
	\end{equation}
	where $\lambda^2_{\mathrm{rms}}\triangleq \sum\nolimits_{i=1}^{N}\lambda^2_{i}/N$, we obtain:
	\begin{IEEEeqnarray}{rCl}
		\IEEEeqnarraymulticol{3}{l}{
		\E\left[\tilde{{\mathbf{v}}}^T_{k_D}{\tilde{{\mathbf{X}}}^T_{k_D}}{\tilde{{\mathbf{X}}}_{k_D}}{\mathbf{\Lambda}}{\tilde{{\mathbf{X}}}^T_{k_D}}{\tilde{{\mathbf{X}}}_{k_D}}\tilde{{\mathbf{v}}}_{k_D}\right]} \nonumber\\
		&=&\E\left[\tilde{{\mathbf{v}}}^T_{k_D}\left.(LN\lambda^2_{\mathrm{rms}}{\mathbf{\Lambda}}+(\nu-1)L^2{\mathbf{\Lambda}}^3\right.)
		\tilde{{\mathbf{v}}}_{k_D}\right]    \nonumber\\
%		&=&LN\lambda^2_{\mathrm{rms}}\E\left[\tilde{{\mathbf{v}}}^T_{k_D}{\mathbf{\Lambda}}\tilde{{\mathbf{v}}}_{k_D}\right] \nonumber\\
%		&&+L^2(\nu-1)\E\left[\tilde{{\mathbf{v}}}^T_{k_D}{\mathbf{\Lambda}}^3
%		\tilde{{\mathbf{v}}}_{k_D}\right].
&=&NL^2\lambda^2_{\mathrm{rms}}\xi_{k_D,{00}} +L^2(\nu-1)\E\left[\tilde{{\mathbf{v}}}^T_{k_D}{\mathbf{\Lambda}}^3
\tilde{{\mathbf{v}}}_{k_D}\right].\label{eq:dblms_mse_expand_7th_simp11}
	\end{IEEEeqnarray}
	%The first term in (\ref{eq:dblms_mse_expand_7th_simp11}) using (\ref{eq:exc_MSE_gen}) can be simplified as $LN\lambda^2_{\mathrm{rms}}\E\left[\tilde{{\mathbf{v}}}^T_{k_D}{\mathbf{\Lambda}}\tilde{{\mathbf{v}}}_{k_D}\right]=NL^2\lambda^2_{\mathrm{rms}}\xi_{k_D,{00}}$, whereas 
	The second term in (\ref{eq:dblms_mse_expand_7th_simp11}) can be approximated as
	\begin{IEEEeqnarray}{rCl}
\IEEEeqnarraymulticol{3}{l}{L^2(\nu-1)\E\left[\tilde{{\mathbf{v}}}^T_{k_D}{\mathbf{\Lambda}}^3
	\tilde{{\mathbf{v}}}_{k_D}\right] } \nonumber \\
		&\approx& L^2(\nu-1)\lambda^2_{\mathrm{rms}}\E\left[\tilde{{\mathbf{v}}}^T_{k_D}{\mathbf{\Lambda}}\tilde{{\mathbf{v}}}_{k_D}\right]\nonumber\\ 
		&=&L^3(\nu-1)\lambda^2_{\mathrm{rms}}\xi_{k_D,00}. \label{eq:dblms_mse_expand_7th_simp_L}
	\end{IEEEeqnarray}
Note that (\ref{eq:dblms_mse_expand_7th_simp_L}) holds under the following assumption:
\begin{equation}\label{assumption1}
\E\left[\tilde{{\mathbf{v}}}^T_{k_D}{\mathbf{\Lambda}}^3
\tilde{{\mathbf{v}}}_{k_D}\right]\cong \E\left[{\mathbf{\Lambda}}^2\right]\E\left[\tilde{{\mathbf{v}}}^T_{k_D}{\mathbf{\Lambda}}
\tilde{{\mathbf{v}}}_{k_D}\right], 
\end{equation}
which is valid under first-order approximation \cite{solo1994adaptive}.

 The middle term in (\ref{eq:dblms_mse_expand}) is then simplified as 
	\begin{equation}\label{eq:dblms_mse_expand_7th_simp_final}
	\rho  (N+(\nu-1)L)L^2\sigma^4\xi_{k_D,00}=\rho PL^2\sigma^4\xi_{{k_D},00},
	\end{equation}
	where
	$P=N+(\nu-1)L$, $\rho =\lambda^2_{\mathrm{rms}}/\sigma^4$ with $\sigma^2=\sum_{i=1}^{N}\lambda_{i}/N$.
	
	%\marginpar{Isn't the "third" and the "second last" the same if there are four terms? "second last" = "penultimate"}
	
	For the second term in (\ref{eq:dblms_mse_expand}), an approximation based on a similar assumption as given in (\ref{assumption1}), gives:
	\begin{IEEEeqnarray}{rCl}\label{eq:dblms_mse_expand_2nd_4th}
		\E\left[\tilde{{\mathbf{v}}}^T_{k_D}{\mathbf{\Lambda}}^2\tilde{{\mathbf{v}}}_{k-1}\right]
		&\approx& \sigma^2L\E\left[\tilde{{\mathbf{v}}}^T_{k_D}{\mathbf{\Lambda}}\tilde{{\mathbf{v}}}_{k-1}\right]\nonumber\\
		&=&\sigma^2L\E\left[\tilde{{\mathbf{v}}}^T_{k-1}{\mathbf{\Lambda}}\tilde{{\mathbf{v}}}_{k_D}\right] \nonumber\\
		&=&\sigma^2L^2\xi_{k-1,0D}.
	\end{IEEEeqnarray}
	
	The second last term in (\ref{eq:dblms_mse_expand}) can be reduced to
	\begin{IEEEeqnarray}{rCl}
\IEEEeqnarraymulticol{3}{l}{ \E{\left[{\mathbf{e}}^T_{{\mathrm{o}},k_D}\tilde{{\mathbf{X}}}_{k_D}\mathbf{\Lambda}\tilde{{\mathbf{X}}}^T_{k_D}{\mathbf{e}}_{{\mathrm{o}},k_D}\right]}}\nonumber \\
		&=&\E\left[\sum\limits_{i=1}^{N}\lambda_{i}\left(\sum\limits_{j=1}^{L}{\tilde{x}}^2_{k_DL+j+i}\right)\right]\E\left[{\mathbf{e}}^T_{{\mathrm{o}},k_D}{\mathbf{e}}_{{\mathrm{o}},k_D}\right]\nonumber\\ 
		&=&LN\lambda^2_{\mathrm{rms}}L\xi_{\mathrm{min}}=\rho L^2 N\sigma^4\xi_{\mathrm{min}}. \label{eq:dblms_mse_expand_8th_simp_final}
	\end{IEEEeqnarray}

Similarly, the last term in (\ref{eq:dblms_mse_expand}) can be simplified to
\begin{equation}\label{eq:noise_inc}
\E\left[{\mathbf{z}}^T_{k_D}{\tilde{{\mathbf{X}}}_{k_D}}{\mathbf{\Lambda}}{\tilde{{\mathbf{X}}}^T_{k_D}}{\mathbf{z}}_{k_D}\right]=\rho L^2 N\sigma^4\sigma^2_z.
\end{equation}

Finally, substituting (\ref{eq:dblms_mse_expand_7th_simp_final})$-$(\ref{eq:dblms_mse_expand_2nd_4th}) into (\ref{eq:dblms_mse_expand}), and after some manipulation, we arrive at
\begin{IEEEeqnarray}{rCl}\label{eq:dblms_mse_expand_combined}
		\xi_{k,00}&\approx&\xi_{{k-1},00}-2{\hat{\hat{\mu}}}L\xi_{k-1,0D}+{\hat{\hat{\mu}}}^2\rho  PL\xi_{k_D,00}\nonumber\\
		&&+\rho LN{\hat{\hat{\mu}}}^2\xi_{\mathrm{min}}+\rho LN{\hat{\hat{\mu}}}^2\sigma^2_z,
\end{IEEEeqnarray}
where $\hat{\hat{\mu}}=\hat{\mu}\sigma^2$.

It is clear from (\ref{eq:dblms_mse_expand_combined}) that $\xi_{k,00}$ can be determined, if the estimate of $\xi_{k-1,0D}$ is known, as $\xi_{\mathrm{min}}$ can be computed from (\ref{eq:min DBMSE}). To estimate $\xi_{k-1,0D}$, we first need to determine any $\xi_{k-1,0s}$, then later we replace $s$ by $D$ without loss of generality. Using (\ref{eq:dblms_transf_simp}) and (\ref{eq:exc_MSE_gen}), we can obtain $\xi_{k-1,0s}$
	\begin{IEEEeqnarray}{-rCl}\label{eq:dblms_mse_expand_cov_term_ana}
L\xi_{k-1,0s}&=&\E\left[\tilde{\mathbf{v}}^T_{k-2}{\mathbf{\Lambda}}\tilde{\mathbf{v}}_{k-s-1}\right]\nonumber \\ 
		&&-\E\left[\hat{\hat{\mu}}\tilde{\mathbf{v}}^T_{k-D-2}\tilde{\mathbf{X}}_{k-D-2}\tilde{\mathbf{X}}^T_{k-D-2}{\mathbf{\Lambda}}\tilde{\mathbf{v}}_{k-s-1}\right]\nonumber\\ 
		&\approx& L\xi_{k-2,0(s-1)}-\hat{\mu}\sigma^2L\E\left[\tilde{\mathbf{v}}^T_{k-D-2}{\mathbf{\Lambda}}\tilde{\mathbf{v}}_{k-s-1}\right].
	\end{IEEEeqnarray}
	Using (\ref{eq:sub2}) and after some simplification of (\ref{eq:dblms_mse_expand_cov_term_ana}), we obtain
	\begin{equation}\label{eq:dblms_mse_expand_simp_ana}
	\xi_{k-1,0s}=\xi_{k-2,0(s-1)}-{\hat{\hat{\mu}}}L\xi_{k-s-1,0(D-s+1)}.
	\end{equation}

For a given delay $D$, it is noticed that (\ref{eq:dblms_mse_expand_simp_ana}) comprises a set of $D$ difference equations for $s\in \{1,2,...,D\}$. Using (\ref{eq:dblms_mse_expand_simp_ana}), a useful relation to determine  $\xi_{k-1,0D}$ is obtained in (\ref{eq: Bf}), see Appendix~\ref{Appendix:B}. Substituting (\ref{eq: Bf}) into (\ref{eq:dblms_mse_expand_combined}) and after some simplification, we get
%It is important to note that the coefficient update equation (\ref{eq:dblms_final}) uses the effective step-size $\hat{\mu}=\mu/L$\marginpar{No, it doesn't.}, hence, the rest of discussion is restrict\marginpar{wrong word?} over it. 	
	%Moreover, it can also be influenced with a non-zero noise, when ${\hat{\hat{\mu}}}$ increases from zero 
	%The convergence behaviour of DBLMS algorithm depends on the root of the denominator in (\ref{eq:dblms_mse_excess_exp_ss}). Under a given $S$, there exists only one real root given by (\ref{eq:dblms_ss_ineq_2}) which governs the stability and convergence of the algorithm. 
	%However, when $2DL$ is much smaller than $\rho (N+2L)$, the effect on $\xi_{\infty,\mathrm{ex}}$ is negligible\marginpar{Why do we consider that then?}.
	%\marginpar{Sentence seems misplaced?}
	\begin{IEEEeqnarray}{rCl}\label{eq:dblms_mse_expand_combined_inter}
		\rho LN{\hat{\hat{\mu}}}^2(\xi_{\mathrm{min}}+\sigma^2_z) & = & \left(2{\hat{\hat{\mu}}}L-\frac{\rho P}{L}\left({\hat{\hat{\mu}}}L\right)^2\right)\xi_{{k_D},00}\nonumber\\
		& &  -2\left({\hat{\hat{\mu}}}L\right)^2D\xi_{k_D,01}+ \xi_{k,00} \nonumber\\
				& &  -\xi_{k-1,00}.
			\end{IEEEeqnarray}
		
To find the excess DBMSE in the steady state, we use the following approximation:
	\begin{equation}\label{eq:dblms_mse_excess_approx}
	\xi_{k,00}\approx\xi_{k-1,00}\approx\xi_{k_D,00}\approx\xi_{k_D,01}\approx \xi_{\infty,00}.
	\end{equation}
 Using (\ref{eq:dblms_mse_excess_approx}), we can simplify (\ref{eq:dblms_mse_expand_combined_inter}) 
	by setting $P=\rho(N+2L)$ ($\because$ $\nu=3$, for Gaussian inputs) as 
	 \begin{equation}\label{eq:dblms_mse_excess_exp_ss_int}	
	2\xi_{\infty,00}-{\hat{\hat{\mu}}}L\left(\frac{\rho (N+2L)}{L}+2D\right)\xi_{\infty,00}=\rho N{\hat{\hat{\mu}}}(\xi_{\mathrm{min}}+\sigma^2_z).	
	\end{equation}
Using (\ref{eq: relation}) and (\ref{eq:dblms_mse_excess_exp_ss_int}), we estimate $\xi_{\infty,\mathrm{ex}}$ in the steady-state as 
	 \begin{equation}\label{eq:dblms_mse_excess_exp_ss}
	 \xi_{\infty,\mathrm{ex}}=\frac{\rho N{\hat{\hat{\mu}}}(\xi_{\mathrm{min}}+\sigma^2_z)}{2-{\hat{\hat{\mu}}}(\rho (N+2L)+2DL)}.
	 \end{equation}
 
From (\ref{eq:dblms_mse_excess_exp_ss}), it can observed that $\xi_{\infty,\mathrm{ex}}$ increases as either $D$ or $L$ or both increases for a fixed step-size. In contrast, if ${\hat{\hat{\mu}}}$ increases from zero, the denominator of (\ref{eq:dblms_mse_excess_exp_ss}) starts decreasing until it becomes zero for a fixed delay $D$ and block size $L$. In consequence, $\xi_{\infty,\mathrm{ex}}$ starts increasing, and could result in algorithm divergence i.e., $\xi_{\infty,\mathrm{ex}}>1$. Thus, an expression for the stability bound and algorithm convergence can be found by setting the denominator of (\ref{eq:dblms_mse_excess_exp_ss}) to $0$. There exists only one real root after solving for $\hat{\hat{\mu}}$ as 	
\begin{equation}\label{eq:dblms_ss_ineq_1}
{\hat{\hat{\mu}}} < \frac{2}{L(\rho (N+2L)/L+2D)}=\frac{2}{\rho N+2(D+\rho)L}.
\end{equation}

Note this is the theoretical limit on the bound for stability and algorithm convergence under the noiseless case. However, due to the involvement of the noise component in the numerator of (\ref{eq:dblms_mse_excess_exp_ss}), its effect on the algorithm convergence slightly reduces the bound. This is explained through the set of simulations in Section \ref{Results}. Without loss of generality, we assume $\sigma^2_z=0$. Using $\hat{\hat{\mu}}=\hat{\mu}\sigma^2$, (\ref{eq:dblms_ss_ineq_1}) in terms of  effective critical step-size $\hat{\mu}_{\mathrm{crit}}$ can be expressed as 
\begin{equation}\label{eq:dblms_ss_ineq_2}
\hat{\mu} < \hat{\mu}_{\mathrm{crit}} \hspace{0.2cm} \mathrm{s.t.} \hspace{0.2cm} \hat{\mu}_{\mathrm{crit}}=\frac{2}{(\rho N+2(D+\rho)L)\sigma^2}.
\end{equation}

By keeping $L=1$ and $D=0$ in (\ref{eq:dblms_ss_ineq_2}) reduces to $\mu_{0,\mathrm{crit}}$ given in (\ref{eq:ss_inequality_lms}), assuming $\rho=1$. Further, $\hat{\mu}_{\mathrm{crit}}$ coincides to $\hat{\mu}_{D, \mathrm{crit}}$ given in (\ref{eq:ss_inequality_dlms}) for $L=1$; while it is slightly smaller $({N+L+1})/({N+2L})$, than $\hat{\mu}_{L, \mathrm{crit}}$ given in (\ref{eq:ss_inequality_Llms}) for $D=0$. The steady-state excess DBMSE from (\ref{eq:dblms_mse_excess_exp_ss}) is then approximated as 
	\begin{equation}\label{eq:dblms_mse_excess_exp1}
	\xi_{\infty,\mathrm{ex}}=\frac{\rho N\hat{\hat\mu} \xi_{\mathrm{min}}}{2-(\rho N+2(D+\rho)L)\hat{\hat\mu}}.   
	\end{equation}

The adaptation accuracy indicates how much coefficient noise is in the steady state, and is quantified by the misadjustment ($M$). It is defined as the ratio of  excess error, $\xi_{\infty,\mathrm{ex}}$ to the minimum MSE, $\xi_{\mathrm{min}}$:
 \begin{IEEEeqnarray}{rCl}\label{eq:misadj_general}
		M\triangleq\frac{\xi_{\infty,\mathrm{ex}}}{\xi_{\mathrm{min}}}=\frac{\rho N\hat{\hat\mu}}{2-(\rho N+2(D+\rho)L)\hat{\hat\mu}}.\label{eq:misadjustment0}
	\end{IEEEeqnarray}

	For given $M$, $N$, $D$, and $L$, the desired $\hat{\mu}$ can be found as
	\begin{equation}
	\hat{\mu}=\frac{2}{(K+2DL)\sigma^2}, \hspace{0.2cm} K=\rho\left(N\left(1+\frac{1}{M}\right)+2L\right).
	\end{equation}

Finally, the optimum effective step size $\hat{\mu}_{\mathrm{opt}}$ can be obtained by differentiating the right side of (\ref{eq:dblms_mse_expand_combined_inter}) with respect to $({\hat{\hat{\mu}}}L)$, setting the result to $0$, and using $\hat{\hat{\mu}}=\hat{\mu}\sigma^2$ as
	\begin{equation}\label{eq:opt_ss_small}
	\hat{\mu}_{\mathrm{opt}}= \frac{1}{(\rho N+2(D+\rho)L)\sigma^2}.  
	\end{equation}

	It is clear from (\ref{eq:opt_ss_small}) that $\hat{\mu}_{\mathrm{opt}}$ is half of $\hat{\mu}_{\mathrm{crit}}$. Therefore, the optimum misadjustment $M_{\mathrm{opt}}$ can be obtained by setting $\hat{\mu}=\hat{\mu}_{\mathrm{opt}}$ in (\ref{eq:misadjustment0}) as
	\begin{equation}\label{eq:misadjustment1}
	M_{\mathrm{opt}}=\frac{\rho N}{(\rho N+2(D+\rho)L)}.
\end{equation}
Clearly, $M_{\mathrm{opt}}$ can be estimated for a given $N, D$, and $L$. For a special case \cite{feuer1985performance}, when all eigen values are equal $\lambda_i=\lambda$ ($i=1,2,\dots,N$) implies $\rho=1$, so (\ref{eq:opt_ss_small}) and (\ref{eq:misadjustment1}) become
\begin{equation}\label{eq:opt_ss_small_rho1}
	\hat{\mu}_{\mathrm{opt}}= \frac{1}{(N+2(D+1)L)\sigma^2}=\frac{1}{(N+2S)\sigma^2}
\end{equation}
and
\begin{equation}\label{eq:misadjustment1_rho1}
M_{\mathrm{opt}}=\frac{N}{(N+2(D+1)L)}=\frac{N}{N+2S},
\end{equation}
respectively. It is clear from (\ref{eq:opt_ss_small_rho1}) and (\ref{eq:misadjustment1_rho1}) that $\hat{\mu}_{\mathrm{opt}}$ and $M_{\mathrm{opt}}$ become fixed for a given choice of $N$ and $S$ ($N$, $D$ and $L$). 

\subsection{Transient-State Analysis} 
	Consider the DBMSE as defined in (\ref{eq:dblms_mse_simp_final}) during the learning process. As the coefficient error vector adapts toward the optimal coefficients ${\mathbf{w}}^{*}$, the error vector ${\mathbf{e}}_{k-D}$ is non-stationary. To model the DBMSE in the transient state, we again compute  $\E\left[\tilde{\mathbf{v}}^T_{k}\mathbf{\Lambda}\tilde{\mathbf{v}}_{k}\right]$, but it requires a different interpretation than excess DBMSE for steady-state (in Appendix~\ref{Appendix:A}). The following DBMSE expression is obtained for the transient state, see Appendix~\ref{Appendix:C}, as
%	\marginpar{check sentence}  
\begin{IEEEeqnarray}{rCl}\label{eq:D3}
		\xi_{k,00}&=&\xi_{{k-1},00}-2{\hat{\hat{\mu}}}L\xi_{k-1,0D}.
	\end{IEEEeqnarray}

	Like (\ref{eq:dblms_mse_expand_combined}), the DBMSE expression in (\ref{eq:D3}) depends on $\xi_{k-1,0D}$. We again follow a set of $D$ difference equations given by (\ref{eq:dblms_mse_expand_simp_ana}) to derive a useful relation between $\xi_{k-1,0D}$ and $\xi_{k-1,00}$ in the transient state, see Appendix~\ref{Appendix:D}. The shorthand notation for (\ref{eq:De}) can be written as:
	\begin{equation}\label{eq:shorthand_matrix_vector}
	{\boldsymbol{\xi}}_{k_D,0}={\mathbf{S}}_{\hat{\hat\mu} L,D}\boldsymbol{\cdot}{\boldsymbol{\xi}_{k_D}},
	\end{equation}
	where ${\mathbf{S}}_{\hat{\hat\mu} L,D}$ is the mapping matrix of size $D \times D$, ${\boldsymbol{\xi}}_{k_D,0}=[\xi_{{k_D},00},0,\dots,0]$ and  ${\boldsymbol{\xi}_{k_D}}=[\xi_{k_D,01},\xi_{k_D,02},\dots,\xi_{k_D,0D}]$. ${\mathbf{S}}_{\hat{\hat\mu} L,D}$ is independent of time instant, and performs mapping from ${\boldsymbol{\xi}_{k_D}}$ to ${\boldsymbol{\xi}_{k_D,0}}$. In scalar form, the relation between $\xi_{k_D,0D}$ and $\xi_{k_D,00}$ can be obtained as	
	\begin{equation}\label{eq:deviation}
		\xi_{k_D,00}=\alpha_{\hat{\hat\mu} L,D} \xi_{k_D,0D},  
		\end{equation}
		where $\alpha_{\hat{\hat\mu}L,D}$ is a slope factor that indicates the change in the transient-state DBMSE as a result of different $D$ and $L$ values. 
		
		Using Cramer's rule, $\alpha_{\hat{\hat\mu}L,D}$ is given by 
		\begin{equation}\label{eq: dev factor}
		\alpha_{\hat{\hat\mu} L,D}=\frac{|{{\mathbf{S}}_{-\hat{\hat\mu} L,(D-1)}|}}{|{\mathbf{S}}_{\hat{\hat\mu} L,D}|},       
		\end{equation}
		where $|\cdot|$ denotes the determinant. After manipulating ${\mathbf{S}}_{\hat{\hat\mu} L,D}$ for both odd and even $D$, as illustrated in Appendix \ref{Appendix:D}, we obtain the recursive expression for $\left|{\mathbf{S}}_{\hat{\hat\mu} L,D}\right|$ as 
		\begin{equation}\label{eq:det_general}
		\left|{\mathbf{S}}_{\hat{\hat\mu} L,D}\right|=\sum\nolimits_{m=0}^{D}(-1)^{\lfloor{m}/{2}\rfloor}\binom{\left\lfloor\frac{D+m}{2}\right\rfloor}{m}\left(\hat{\hat\mu} L\right)^m,
		\end{equation}
		where $\lfloor \cdot \rfloor$ denotes the floor function, and $\binom{\cdot}{\cdot}$ denotes the binomial coefficients. $\left|{\mathbf{S}}_{-\hat{\hat\mu} L,(D-1)}\right|$ can be obtained from (\ref{eq:det_general}) by replacing $\hat{\hat\mu} L$ with $-\hat{\hat\mu} L$ and $D$ with $D-1$.
		
\begin{figure}
		\centering
		\includegraphics[width=1\linewidth]{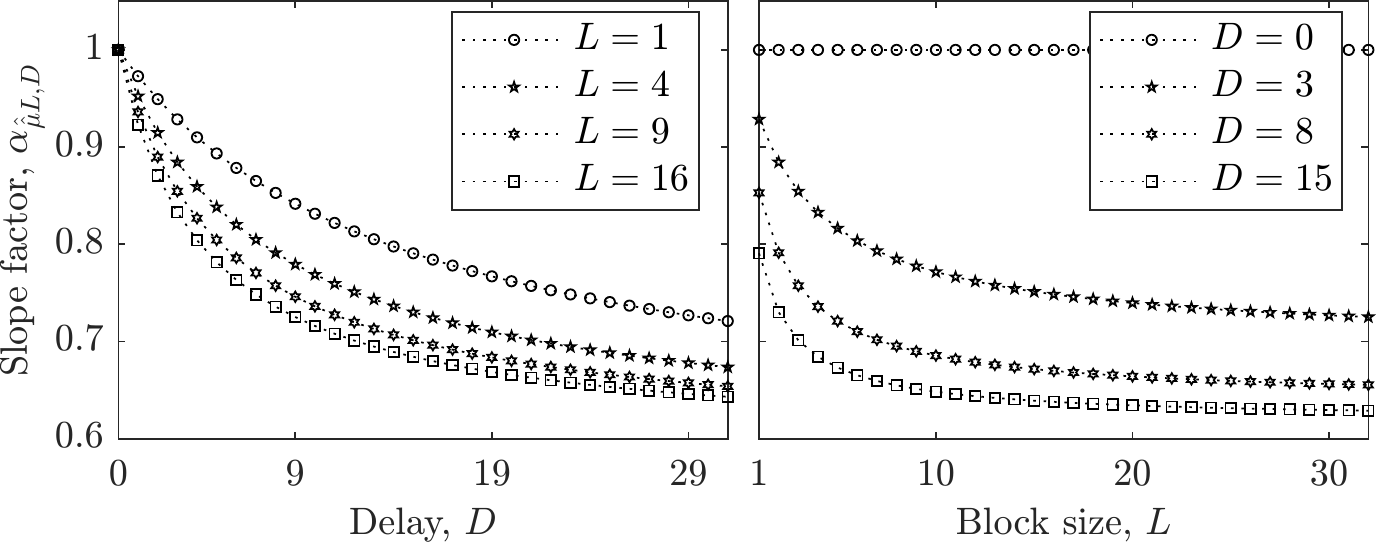}
		\caption{Behaviour of $\alpha_{\hat{\hat\mu}L,D}$ for $N=32$ as a function of $D$ for $L \in \{1, 4, 9, 16\}$ (left), and $L$ for $D \in \{0, 3, 8, 15\}$ (right).}
		\label{fig_alpha}
	\end{figure}

		The behaviours of $\alpha_{\hat{\hat\mu}L,D}$ for $N=32$ and $\hat{\mu}=\hat{\mu}_{\mathrm{opt}}$ with $D$ for some $L \in \{1, 4, 9, 16\}$, and with $L$ for some $D \in \{0, 3, 8, 15\}$ are shown in left and right of Fig.~\ref{fig_alpha}, respectively. From the curves, it is clear that  $\alpha_{\hat{\hat\mu}L,D}$ decreases for both increasing $D$ and $L$ values. The rate of decrease in $\alpha_{\hat{\hat\mu}L,D}$ for a given $L$ with different $D$ values is slower than the rate of decrease in $\alpha_{\hat{\hat\mu}L,D}$ for a given $D$ with different $L$ values. It is interesting to note that $\alpha_{\hat{\hat\mu}L,D}$ saturates slower for a larger $D$ with a given $L$, while it saturates faster for a smaller $L$ with a given $D$. Using (\ref{eq:deviation}), we can re-write (\ref{eq:D3}) as
		\begin{IEEEeqnarray}{rCl}\label{eq:exc_DBMSE_transient}
			\xi_{k,00}&=&\xi_{{k-1},00}-2{\hat{\hat{\mu}}}L\alpha_{\hat{\hat\mu} L,D}\xi_{{k-1},00}\nonumber\\
			&=&\left(1-2{\hat{\hat{\mu}}}L\alpha_{\hat{\hat\mu} L,D}\right)^k\xi_{0,00}.
		\end{IEEEeqnarray}
		Clearly, DBMSE in the transient state decays geometrically with a ratio $\left(1-2{\hat{\hat{\mu}}}L\alpha_{\hat{\hat\mu} L,D}\right)$. 
		%\begin{equation}\label{eq:D5}
		%r_{{\hat{\hat\mu} }L,D}=\left(1-2{\hat{\hat{\mu}}}L\alpha_{\hat{\hat\mu} L,D}\right).  
		%\end{equation}
		Like $\xi_{k,00}$ in (\ref{eq:exc_DBMSE_transient}), the actual $\xi_{k}$ also decays geometrically, as per (\ref{eq:dblms_mse_simp_final}), when $\hat{\hat\mu}$ is appropriately chosen for DBLMS convergence. Observably, it can be seen that $\displaystyle{\lim_{k \to \infty}} \xi_k=\xi_{\mathrm{min}}$ which then follows 
	\begin{equation}\label{eq:D6}
	\displaystyle{\lim_{k \to \infty}} \left(1-2{\hat{\hat{\mu}}}L\alpha_{\hat{\hat\mu} L,D}\right)^k=0 .    
	\end{equation}

	To follow the decay, we define a time constant of DBMSE convergence for an exponential envelope to fit on the above geometric sequence.  Its value should be $2$ times that of the time constant needed for the coefficients converging toward their optimal values ${\mathbf{w}}^{*}$ \cite{clark1981block1}. This is because the excess DBMSE in (\ref{eq: excess_MSE_def}) involves two coefficient vectors due to second-order statistics. Thus, we can equate the geometric ratio to the exponential envelope:
%	\marginpar{Do we assume this? Yes, we are considering order statistics. Eq. (4) involves first order statistics and does not give the guarantee to converge. I agree with you that it is not an assumption. I am citing reference [28] for justification}
	
    \begin{equation}\label{eq:D7}
	\left(1-2{\hat{\hat{\mu}}}L\alpha_{\hat{\hat\mu} L,D}\right)=e^{-\frac{1}{2\tau}}=1-\frac{1}{2\tau}+\frac{1}{2!(2\tau)^2} \cdots.
	\end{equation}

	For the case of a slow adaptation, which is of most general interest, $\tau$ is usually large, so that (\ref{eq:D7}) can be approximated by the first two terms of the Taylor series expansion which results in 
	\begin{equation}\label{eq:D8}
	\tau = \frac{1}{4{\hat{\hat{\mu}}}L\alpha_{\hat{\hat\mu} L,D}} \text{ with } \tau_{\mathrm{opt}} = \frac{N+ 2S}{4\alpha_{\hat{\hat\mu}_\mathrm{opt} L,D}}  ,  
	\end{equation}    
	where $\tau_{\mathrm{opt}}$ is the time constant at $\hat{\hat{\mu}}=\hat{\hat{\mu}}_{\mathrm{opt}} = \hat{\mu}_{\mathrm{opt}}\sigma^2$.

	It can be noted that $\tau$ is of the same form as that of the non-pipelined non-block method ($D=0$, $L=1$) of steepest descent \cite{clark1981block1}. Furthermore, it only depends on the behaviour of $\alpha_{\hat{\hat\mu} L, D}$ for a given $\hat{\hat{\mu}}$. 
	
	The adaptation speed of the DBLMS ADF can be determined from the transient decay (or slope) of the MSE curve. According to (\ref{eq:D7}), the exponential envelope decay involving $\tau$ decides the adaptation speed. Thus, the slope in dB/sample can be defined as
	\begin{equation}\label{eq:speed}
		\mathrm{Slope} = -\frac{10\log_{10}(e)}{2\tau}\text{ with }\mathrm{Slope}_{\mathrm{opt}} \approx -8.68 \frac{\alpha_{\hat{\hat\mu}_\mathrm{opt} L,D}}{N+2S},
	\end{equation}
	%where (\ref{eq:tau_same_speedup}) is used for $\mathrm{Slope}_{\mathrm{opt}}$. %Clearly, the adaptation speed is a function of ${\tau_{\mathrm{opt}}}$, which solely depends on the nature of $\alpha_{\hat{\hat\mu}L,D}$ under a given speedup $S=(D+1)L$.
	where $\mathrm{Slope}_{\mathrm{opt}}$ is the slope approximation using $\tau_{\mathrm{opt}}$.
	%To make the units of non-block and block methods consistent, we keep the actual step-size in (\ref{eq:D8}) using ${\hat{\hat{\mu}}}={\hat{\mu}}\sigma^2$ and ${\hat{\mu}}=\mu/L$ as 
%	\begin{equation}\label{eq:time_constant}
%	\tau=L\hat\tau=\frac{1}{4{\hat{\mu}\sigma^2}\alpha_{\hat{\hat\mu} L,D}},
%	\end{equation}
	%where ${\hat{\hat{\mu}}}={\hat{\mu}}\sigma^2$. 
%	It is clear from (\ref{eq:time_constant}) that 
	
	%In terms of speedup $S$ and $\hat{\mu}=\hat{\mu}_{\mathrm{opt}}$, a simplified expression for optimum time constant $\tau_{\mathrm{opt}}$ can be obtained using (\ref{eq: opt_ss_small_rho1}) and (\ref{eq:time_constant}) as
	%\begin{equation}\label{eq:tau_same_speedup}
	%\tau_{\mathrm{opt}}=\frac{N+2S}{4\alpha_{\hat{\hat\mu} L,D}}.    
	%\end{equation}
	%Unlike (\ref{eq: opt_ss_small_rho1}) and (\ref{eq:misadjustment1_rho1}), $\tau_{\mathrm{opt}}$ depends on the behavior of $\alpha_{\hat{\hat\mu} L,D}$, as it is not completely explicit in terms of $S$. 

%The block diagram of system identification problem based on DBLMS ADF for an unknown plant with coefficients ${\mathbf{w}}^o$ is shown in Fig. \ref{fig:system_identification}. 
 
\section{Results and Discussion}\label{Results}
This section presents the application of ADF to system identification problem to corroborate the proposed model and illustrate the properties of the DBLMS algorithm. Validations of analytical step-size bound $\hat{\mu}_{\mathrm{crit}}$ (\ref{eq:dblms_ss_ineq_2}), adaptation accuracy in terms of misadjustment, $M_{\mathrm{opt}}$ (\ref{eq:misadjustment1_rho1}), and adaptation speed using (\ref{eq:D8}) are included through extensive simulations. The simulations are carried out with white Gaussian data which keeps sufficient information for the estimated results and provides reliable design guidelines \cite{key-0}.

Note that the initial coefficient vector is zero in all cases. The Monte Carlo simulations are performed by averaging 500 ensemble trials. For the considered examples, the coefficients ${\mathbf{w}}^o$ of unknown plant are normalized to satisfy ${{\mathbf{w}}^o}^T{\mathbf{w}}^o=1$. The order of the unknown plant and the taps of DBLMS ADF are considered to be the same size. In all the simulations, white Gaussian input ${\mathbf{X}}_k$ consisting of independent samples with zero mean and unit variance is used. While the desired input $d_n$ is obtained by contaminating the output of the unknown plant with white Gaussian noise $z_n$ of strength -$60$ dB. 
%In all the simulations, 32-taps DBLMS ADF is considered, unless stated otherwise. 
%\begin{figure}
%	\centering
%	\includegraphics[width=0.80\linewidth]{sys_ident.pdf}
%	\caption{Block diagram of system identification problem with DBLMS ADF.} 
%	\label{fig:system_identification}
%\end{figure}
%\begin{figure}
%	\centering
%	\includegraphics[width=1\linewidth]{stability_results.pdf}
%	\caption{Critical step-size $\hat{\mu}_{\textrm{crit}}$ with white Gaussian noise. (a) vs $D$ for $N=4$ and $L \in \{1 ,4\}$. (b) vs $L$ for $N=4$ and $D \in \{0 ,3\}$. (c) vs $D$ for $N=32$ and $L \in \{1, 4, 32\}$. (d) vs $L$ for $N=32$ with $D \in \{0, 3, 31\}$. Analytical: normal lines. Simulated (ensemble): dashed-dotted lines. Simulated (worst case): dotted lines.}
%	\label{fig5}
%\end{figure}

%The ensemble simulated stability bound is always higher than the worst-case due to averaging. %Based on this strategy, we are able obtain two simulated stability bounds, that is, one  by averaging the ensemble trials, and other is standalone worst-case. 
%While in the case of $N=32$, the analytical stability bound matches well with both simulated (ensemble) and simulated (worst) except for low $D$ in case of DLMS and low $L$ in case of BLMS, as shown in Fig.~\ref{fig5}(c)-(d). Interestingly, the matching of analytical and simulated is quite well for the DBLMS. 

\subsection{Stability Analysis}
\begin{figure}
	\centering
	\includegraphics[width=1\linewidth]{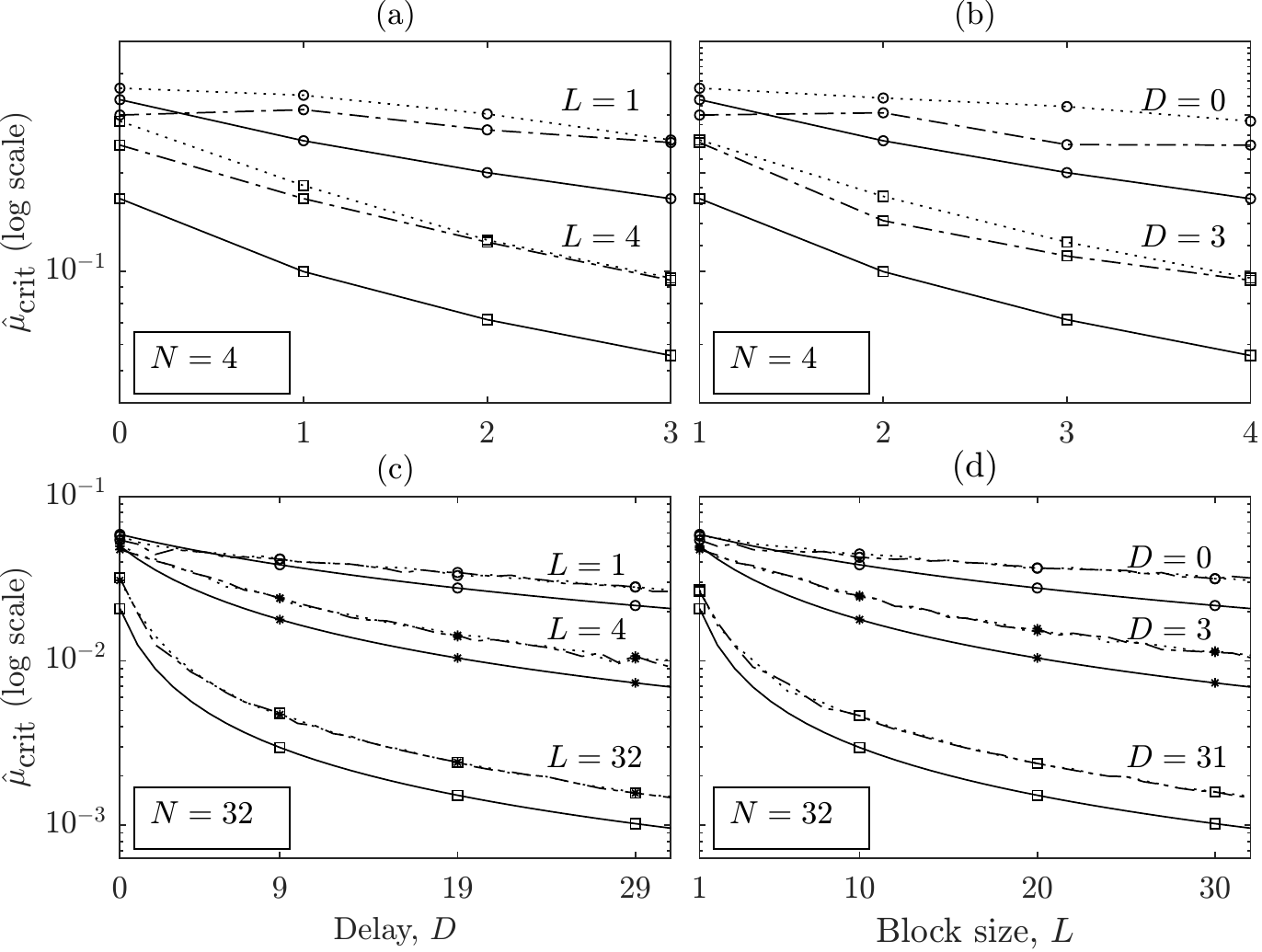}
	\caption{Critical step-size $\hat{\mu}_{\textrm{crit}}$. (a) vs $D$ for $N=4$ and $L \in \{1 ,4\}$. (b) vs $L$ for $N=4$ and $D \in \{0 ,3\}$. (c) vs $D$ for $N=32$ and $L \in \{1, 4, 32\}$. (d) vs $L$ for $N=32$ with $D \in \{0, 3, 31\}$. Analytical: normal lines. Simulated (noiseless): dotted lines. Simulated (with noise of -60 dB): dashed-dotted lines.}
	\label{fig5}
\end{figure}

The analytical prediction and simulation results to validate the stability bound given by (\ref{eq:dblms_ss_ineq_2}) are shown in Fig.~\ref{fig5}(a)--(d). The simulation results are obtained by incrementing the step size in $0.0265\hat{\mu}_{\mathrm{crit}}$ divisions, while simultaneously checking the magnitude of any steady-state MSE point $\geq 1$, for algorithm divergence. Precisely, at every incremental step, we check if any sample of steady-state MSE crosses the unity. This acts as a starting point to find out the critical stability. Two different scenarios corresponding to small taps, e.g., $N=4$, and large taps, e.g., $N=32$ with different delays and block sizes are considered. This is to demonstrate that our analytical results apply to any tap size, delay, and block size.

The theoretical bound as given in (\ref{eq:dblms_ss_ineq_2}), matches fairly well the extensive simulations for both the taps and different delays and block sizes. The analytical step-size bound for $N=4$ is slightly overestimated by the simulated bound for different delays and block sizes with a noiseless case, as shown in Fig.~\ref{fig5}(a)--(b). However, the simulated bound slightly shifts down with $-60$ dB Gaussian noise. This is because the impact of increasing the step-size towards bound on excess MSE given in (\ref{eq:dblms_mse_excess_exp_ss}) will be slightly favoured through a noise. As a consequence, it causes a slight mismatch between the simulated and analytical results, especially at lower delays and  block sizes. Similar arguments apply to simulated and analytical results for $N=32$, as shown in Fig.~\ref{fig5}(c)-(d).

These studies have not been considered in the prior works for DLMS and BLMS algorithms \cite{long1992corrections,ernst1995analysis, feuer1985performance, lee2012step}. Among \cite{long1992corrections,ernst1995analysis, feuer1985performance, lee2012step}, only \cite{ernst1995analysis} had described a partial strategy for simulating the bound with random binary inputs. Similarly, leaky-DLMS \cite{tobias2004leaky} simulated the bound under a noiseless scenario for $N=32$. They showed that the simulated stability bound lies always below the analytical stability bound. From this observation, it is believed that the considered strategy for the simulated bound is well explored and accurate enough to validate the analytical bound given in (\ref{eq:dblms_ss_ineq_2}).    

Furthermore, one can observe the mismatches between the analytical and simulated stability bound results, the reasons are described as follows: (1) the approximations used in (\ref{eq:dblms_mse_expand_7th_simp_L}), (\ref{assumption1}), (\ref{eq:dblms_mse_expand_2nd_4th}), (\ref{eq:dblms_mse_expand_cov_term_ana}) and (\ref{eq:dblms_mse_excess_approx}); and (2) the independent assumption \cite{mazo1979independence, tobias2004leaky}. In addition, some approximations are considered such as equal eigenvalues like the one given in (\ref{eq:opt_ss_small_rho1}) and (\ref{eq:misadjustment1_rho1}). Even so, the simulation results presented in Fig.~\ref{fig5} illustrate that the stability bound as given in (\ref{eq:dblms_ss_ineq_2}) is a valid approximation.

\subsection{Adaptation Accuracy}
The analytical misadjustment $M_{\mathrm{opt}}$ as derived in (\ref{eq:misadjustment1_rho1}) is also verified through simulations. We consider two different scenarios which are more practical, namely, the one with speedup is the same as the number of taps, i.e., $S=N$; and the other with speedup is less than the number of taps, i.e., $S<N$. For $S>N$, it violates the assumption (\ref{assumption1}), especially with large $D$ values, thereby making the proposed model less accurate \cite{solo1994adaptive}. 
%Under any speedup $S=(D+1)L$, it is possible to select different $D$ and $L$ values, for instance, $S=4$ allows to select different $D$ and $L$ from an ordered pair $(D, L) \in \{(0,4), (1,2), (3,1)\}$. 

\begin{figure}
	\centering
	\includegraphics[width=1\linewidth]{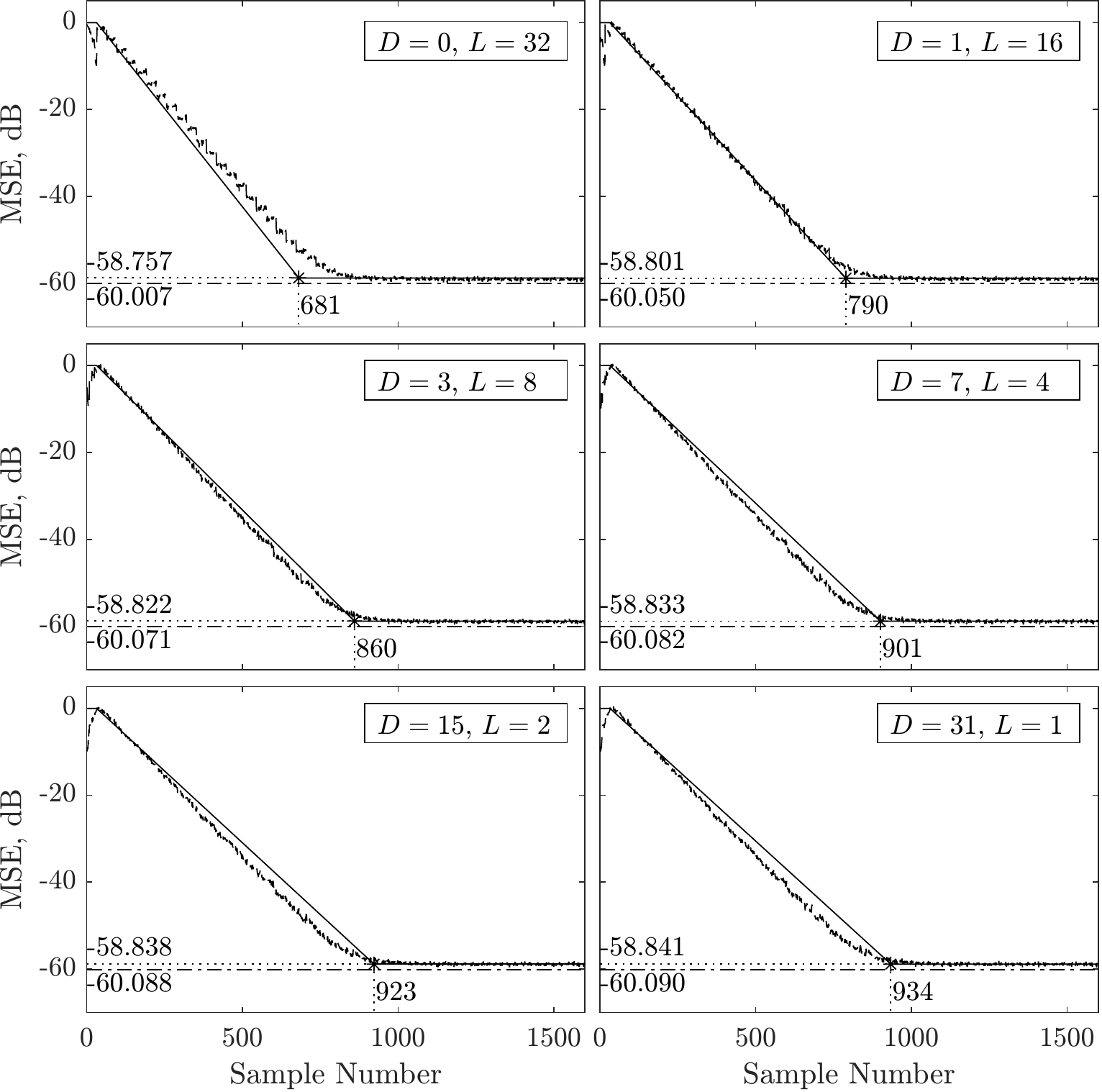}
	\caption{Convergence curves of DBLMS ADF for $S=N$ with $N=32$ and $(D,L) \in \{(0,32),(1,16), (3,8), (7,4), (15,2), (31,1)\}$, where the analytical predictions given in (\ref{eq:opt_ss_small_rho1}) and (\ref{eq:misadjustment1_rho1}) are indicated by a solid line; simulations output are indicated by a dashed line; theoretical minimum MSE $\xi_{\mathrm{min}}$ given in (\ref{eq:min DBMSE}) are indicated by a dash-dotted line; the point of convergence is indicated by $*$ with coordinates indicated by dotted lines.} 
	\label{fig:same_speedup}
\end{figure}

To begin with, we first consider $S=N$ with a system identification example to discuss the proposed model, and we later discuss $S<N$. The possible $D$ and $L$ values for a DBLMS ADF with $N=32$ and $S=32$ can be selected from an ordered pair $(D, L)\in \{(0, 32), (1, 16), (3, 8), (7, 4), (15, 2), (31, 1)\}$. Both the simulated and analytical convergence curves are shown in Fig.~\ref{fig:same_speedup}. In all the cases, the step size $\hat{\mu}_{\mathrm{opt}}$ given in (\ref{eq:opt_ss_small_rho1}) is used. The analytical $\xi_{\mathrm{min}}$ corresponding to the Wiener solution given in (\ref{eq:min DBMSE}) is estimated. It can be observed that $\xi_{\mathrm{min}}$ shown in Fig.~\ref{fig:same_speedup} reduces slightly for higher $D$ and lower $L$ values, for instance, $\xi_{\mathrm{min}}=-60.007$ dB for $D=0,L=32$ and $\xi_{\mathrm{min}}=-60.090$ dB for $D=31,L=1$. This slight reduction is due to the increased cross-correlation between ${\mathbf{X}}_{k-D}$ and ${\mathbf{d}}_{k-D}$ for lower $L$ and higher $D$ values, in accordance with  (\ref{eq:min DBMSE}). The analytical $\xi_{\mathrm{\infty,ex}}$ is then estimated by the analytical $\xi_{\mathrm{min}}$ using the misadjustment $M_{\mathrm{opt}}$ given in (\ref{eq:misadjustment1_rho1}). It is interesting to note that $\xi_{\mathrm{\infty,ex}}$ follows the same trend as that of $\xi_{\mathrm{min}}$ since $M_{\mathrm{opt}}$ is constant irrespective of $D$ and $L$ values. Moreover, these correspond to approximately the same steady-state MSE.

For evaluation, the estima and simulated $M_{\mathrm{opt}}$ given in (\ref{eq:misadjustment1_rho1}) for different possible $D$ and $L$ values are listed in Table~\ref{tab3}. To determine the simulated $M_{\mathrm{opt}}$, only simulated $\xi_{\mathrm{\infty,ex}}$ is to be estimated, as $\xi_{\mathrm{min}}$ is known using (\ref{eq:min DBMSE}). By averaging the steady-state MSE samples, the simulated $\xi_{\mathrm{\infty,ex}}$ is estimated. %The corresponding simulated $M_{\mathrm{opt}}$ for possible $D$ and $L$ values are listed in Table~\ref{tab3}.
It can be seen that the simulated $M_{\mathrm{opt}}$ increases for lower $L$ and higher $D$ values and matches fairly well to the proposed model. For the chosen $D$ and $L$ values, the analytical point of convergence (i.e., the intersection of analytical exponential decay with analytical excess error steady-state MSE) approximates fairly to the simulated point of convergence. The simulations show that a higher $D$ and a lower $L$ increases the misadjustment, and vice-versa. 

\begin{table}
	\centering
	\caption{Estimated and Simulated Optimal Misadjustment, $M_{\mathrm{opt}}$, (\ref{eq:misadjustment1_rho1}) and $\mathrm{Slope}_{\mathrm{opt}}$   (\ref{eq:speed}) with $S=N$ and $S<N$.} \label{tab3}
	\begin{threeparttable}
		\begin{tabular}{l|l|c|c|c|c}
			%\hline
			&  &   \multicolumn{2}{c|}{
				%Opt. misadjustment, 
				$M_{\mathrm{opt}}$}   &\multicolumn{2}{c}{
				%Opt. time constant, 
				$\mathrm{Slope}_{\mathrm{opt}}$$^\dagger$}                         \\ \cline{3-6} 		
			
			$N$&  $S, D, L$ & Est. (\ref{eq:misadjustment1_rho1}) & $^{\ast}$Simulated & Est. (\ref{eq:speed}) & $^{\ast\ast}$Simulated \\ \cline{1-6}

			32	& 32, 0, 32               & 0.333                    &  0.255  &      -0.091               & -0.078    \\ %\cline{2-6}
			& 32, 1, 16                & 0.333                    & 0.294 &            -0.078         &  -0.076   \\ %\cline{2-6}
			& 32, 3, 8                & 0.333                    &  0.318  &            -0.071         & -0.075   \\ %\cline{2-6}
			& 32, 7, 4                 & 0.333                    & 0.334 &              -0.068       &  -0.073   \\ %\cline{2-6}
			& 32, 15, 2                & 0.333                    &  0.344 &              -0.066       &  -0.072  \\ %\cline{2-6}
			& 32, 31, 1                & 0.333                    &  0.349 &               -0.065      &  -0.071   \\ \hline
			
			16	& 12, 0, 12              &             0.400        & 0.265    &    -0.217                 & -0.173     \\ %\cline{2-6}
			&12, 1, 6              &  0.400                    &  0.340   &           -0.189          & -0.162   \\ %\cline{2-6}
			& 12, 2, 4               & 0.400                    & 0.364  &             -0.179       &  -0.156    \\ %\cline{2-6}
			& 12, 3, 3                 &             0.400        & 0.367 &              -0.174       &  -0.154   \\ %\cline{2-6}
			&12, 5, 2                &  0.400                    &  0.381 &                -0.169     &  -0.152  \\ %\cline{2-6}
			& 12, 11, 1               & 0.400                    & 0.389  &                -0.165   &   -0.149   \\ \hline

		\end{tabular}
		\begin{tablenotes}
			\item[$\dagger$]: Unit of slope is dB/sample, $^{\ast}$: Calculated by (\ref{eq:misadj_general}) with $\xi_{\infty,\mathrm{ex}}$ is determined by averaging steady-state samples, and $\xi_{\mathrm{min}}$ is determined by (\ref{eq:min DBMSE}), $^{\ast\ast}$: Calculated by two points $(a_1,b_1)$ and $(a_2,b_2)$ in the linear region of MSE curve using the relation $(b_2-b_1)/(a_2-a_1)$. 
			%\item[$\dagger$]  Unit of $\tau_{\mathrm{opt}}$ is the dB per sample $-10 \log_{10}(e) = -4.34$ dB. 
		\end{tablenotes}
	\end{threeparttable}
\end{table}

\begin{figure}
	\centering
	\includegraphics[width=1\linewidth]{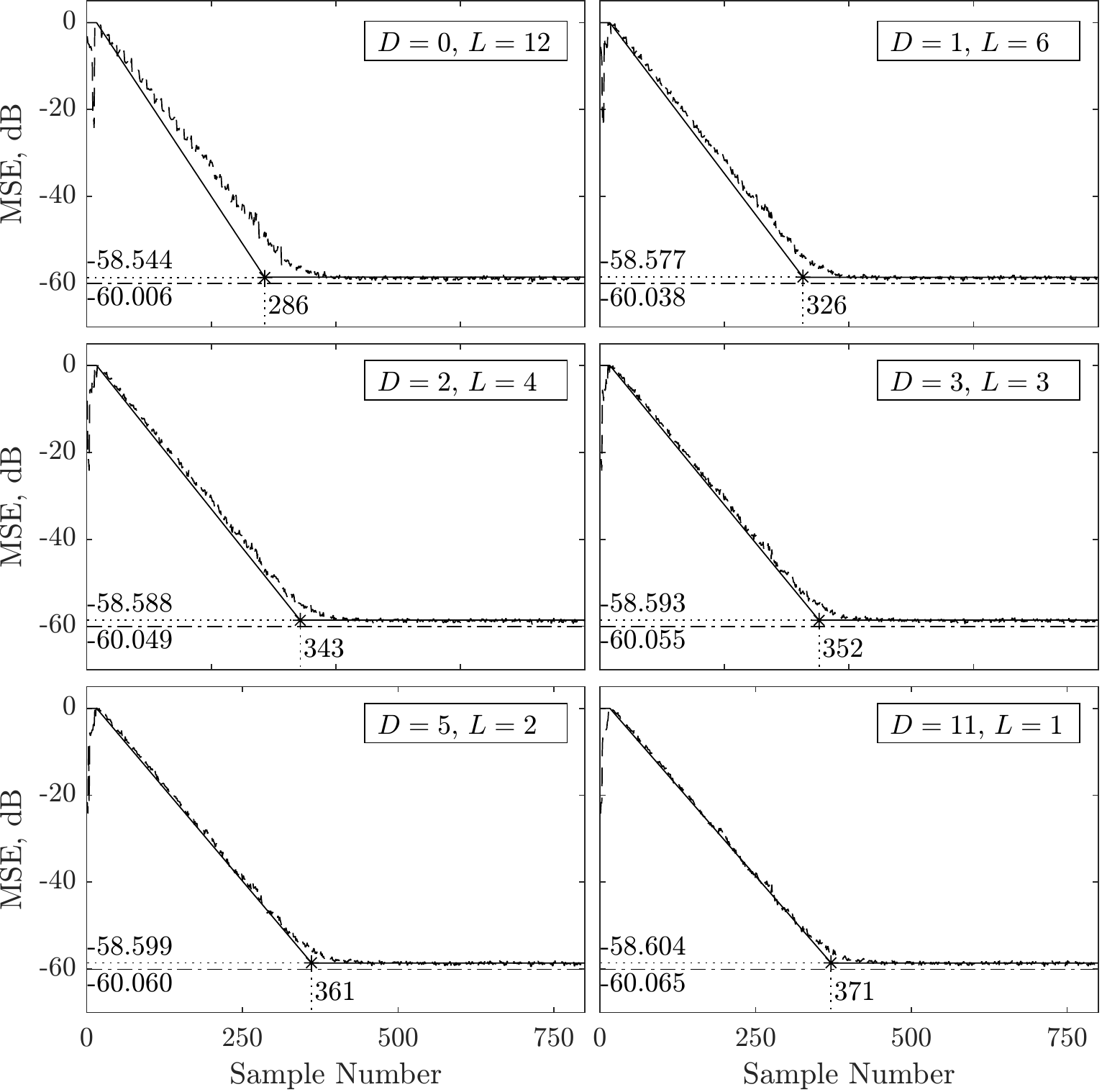}
	\caption{Convergence curves of DBLMS ADF for $S<N$ with $N=16$ and $(D,L) \in \{(0,12), (1,6), (2,4), (3,3), (5,2), (11,1)\}$, where the meaning of different linestyles for analytical predictions are same as those in Fig.~\ref{fig:same_speedup}.} 
	\label{fig:same_speedup1}
\end{figure}

Likewise, $S<N$ is considered, however, with $N=16$ and $S=12$. The selection of different $N$ is to reinforce that the proposed model holds for various $N$ and $S$. In this case, the possible $D$ and $L$ values can be selected from an ordered pair $(D, L) \in \{(0,12), (1,6), (2,4), (3,3), (5,2), (11,1)\}$. Both the simulated and analytical MSE curves of the DBLMS ADF are shown in Fig.~\ref{fig:same_speedup1}. 

The same arguments are also valid for $S<N$ as is the case of $S=N$. However, both analytical and simulated $\xi_{\mathrm{min}}$ and $\xi_{\mathrm{\infty,ex}}$ reduce a bit, thereby increasing both analytical and simulated $M_{\mathrm{opt}}$. This can also be understood from the fact that when $S<N$, $\hat{\mu}_{\mathrm{opt}}$ given in (\ref{eq:opt_ss_small_rho1}) is increased by $3N/(N+2S)$ times as compared to  $S=N$ case. As a consequence, the misadjustment is reduced, as per (\ref{eq:misadjustment0}). The analytical and simulated $M_{\mathrm{opt}}$ values so obtained are listed in Table~\ref{tab3}.    

\subsection{Adaptation Speed}
 To validate (\ref{eq:speed}), we reconsider the $S=N$ and $S<N$ scenarios. In the transient state, both the simulated and analytical convergence curves for $S=N$ and $S<N$ are shown in Figs.~\ref{fig:same_speedup}~and~\ref{fig:same_speedup1}, respectively.

From Fig.~\ref{fig_alpha}, it can be observed that $\alpha_{\hat{\hat\mu}L,D}$ reduces more quickly for a higher $L$ and a lower $D$, and vice-versa. This implies ${\tau_{\mathrm{opt}}}$ is smaller for a higher $L$ and a lower $D$, which brings the faster convergence, and vice-versa. This can be explained by using the analytical point of convergence i.e., the number of samples taken by DBLMS ADF for possible $D$ and $L$ values to reach the steady-state. For given a speedup $S\leq N$, the selection of a higher $D$ and a lower $L$ shifts the point of convergence towards the right (shown in Figs. ~\ref{fig:same_speedup} and ~\ref{fig:same_speedup1}), i.e., reduces the convergence rate, and vice-versa. 

For instance, the number of samples for $S=N$ with $N=32$ is $681$ for $(D,L) = (0, 31)$, and is $934$ for $(D, L) = (31, 1)$, as shown in Fig. ~\ref{fig:same_speedup}. In contrast, for any $S$, especially under $S<N$, the step-size $\hat\mu_{\mathrm{opt}}$ is increased, therefore, the adaptation speed becomes higher as compared to $S=N$ case. A more appropriate measure to determine the adaptation speed is the slope of the MSE curve in the transient state, as discussed above. For better validation of adaptation speed, both the analytical and simulated slope under $S=N$ and $S<N$ scenarios are also listed in Table~\ref{tab3}. It can be observed that the simulated slope is well captured by the proposed analytical model for different possible $(D, L)$ values.

\begin{figure}
	\centering
	\includegraphics[scale=0.62]{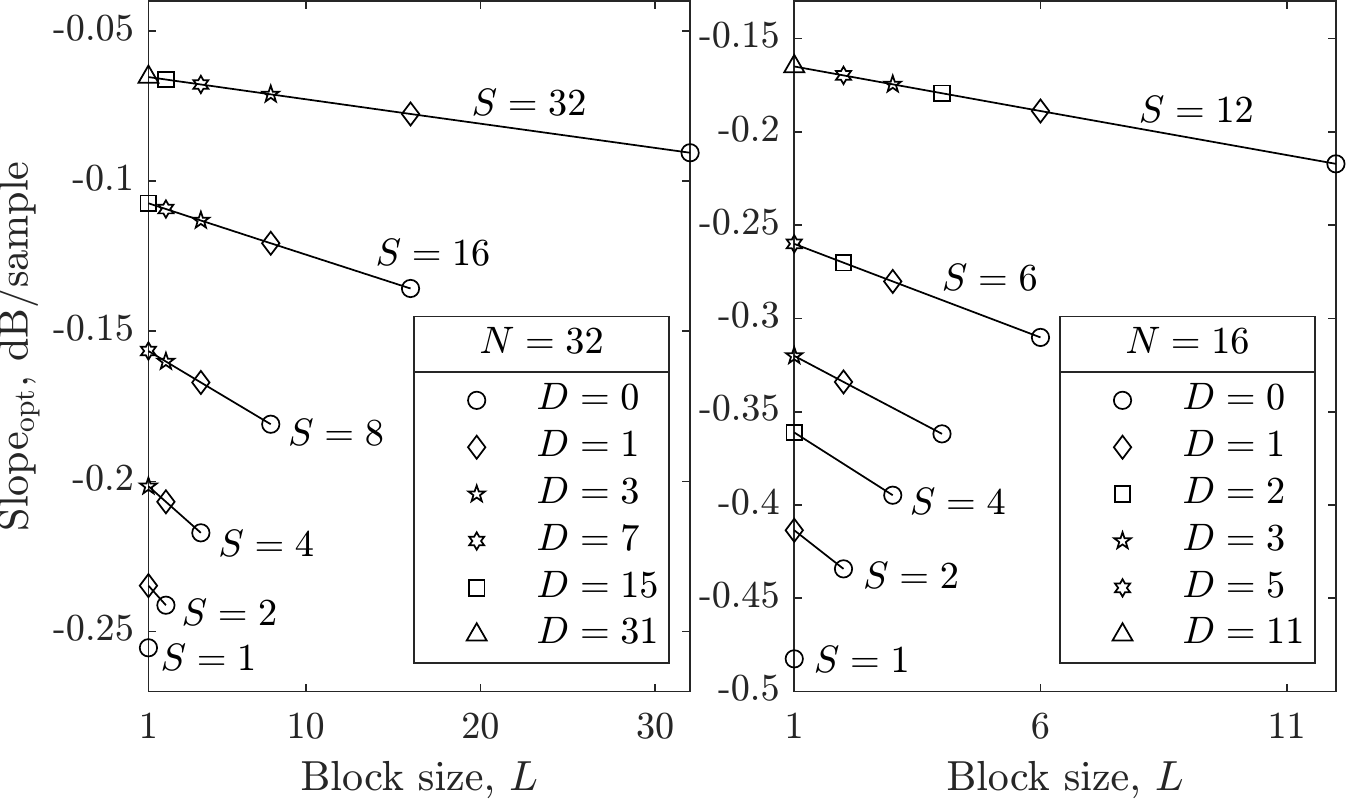}
	\caption{Slope variation with $S\leq N$ for $N=32$ (left) and $N=16$ (right).}\label{fig:tau_variation}
\end{figure}

In the above discussion, we have considered the single speedup $S\leq N$, while it can take any value in practice. To have a better insight into the adaptation speed with different speedups, the slope for any speedup $S$ in power (and/or non-power) of twos for different $D$ and $L$ values is illustrated in Fig.~\ref{fig:tau_variation}. It is found that for a given $S$ and $L$, the increase in $D$ also increases the $\tau$, thereby reducing the slope. However, choosing a higher $L$ can converge the DBLMS ADF faster. Noticeably, for any $S$ in the power of twos will impose the selection of $D$ from the odd values, as shown for $N=32$ in Fig. ~\ref{fig:tau_variation}. In contrast, when $S$ is a non-power of two (or a composite number) will allow the selection of $D$ from both even and odd values, as shown for $N=16$ in Fig. ~\ref{fig:tau_variation}.

%The analytical rate of decay is obtained by the exponential envelope given in (\ref{eq:D7}) with the time constant (\ref{eq:tau_same_speedup}). 
%\marginpar{But this holds for general $S$, not just $S < N$? Yes, I agree with you}

%\subsection{Comparison with Existing DLMS and BLMS models}
%To compare the proposed model of DBLMS algorithm with those of existing step-size bounds given in (\ref{eq:ss_inequality_dlms}) and (\ref{eq:ss_inequality_Llms}), we keep $L=1$ for DLMS algorithm and $D=0$ for BLMS algorithm in (\ref{eq:dblms_ss_ineq_2}) to obtain their step-size bounds as
%\begin{equation}
%\mu^{\mathrm{P}}_{D,\mathrm{crit}}=\frac{2}{(N+2+2D)\sigma^2} \text{ and }  \mu^{\mathrm{P}}_{L,\mathrm{crit}}=\frac{2L}{(N+2L)\sigma^2}.     
%\end{equation}
%Observably, the obtained step-size bound $\mu^{\mathrm{P}}_{D,\mathrm{crit}}$ for the DLMS algorithm coincides with (\ref{eq:ss_inequality_dlms}); and the obtained step-size bound $\mu^{\mathrm{P}}_{L,\mathrm{crit}}$ for the BLMS algorithm is slightly smaller, $({N+L+1})/({N+2L})$, than (\ref{eq:ss_inequality_Llms}). In addition, setting $L=1$ and $D=0$ gives $\mu_{0,\mathrm{crit}}$ of the LMS algorithm given in (\ref{eq:ss_inequality_lms}). 

%Similarly, for a stable conventional BLMS \cite{clark1981block}, the step-size bound is slightly reduced $({N+2})/({N+2(D+1)L})$. This is the first time to show an explicit step-size bound that has both delay and block size parameters for the validation of stability and algorithm convergence. Interestingly, the revelation of this not-so-restrictive bound, and would allow to development of efficient delayed block adaptive filters for high throughput applications.

\section{Conclusion}
A stochastic analysis of the LMS algorithm with delayed block coefficient adaptation which is very useful to obtain high sample rates adaptive filter has been presented. The delayed block adaptive filtering procedure has been considered in which the coefficients are adjusted once per block of delayed data in the LMS sense. Then, an analysis has been carried out to calculate step-size bound and adaptation accuracy. Subsequently, a measure of the adaptation speed in the transient state for a given delay and block size has been provided.

This is the first time to show an explicit step-size bound that has both delay and block size parameters for the validation of stability and algorithm convergence. Interestingly, the revelation of this not-so-restrictive bound, and would allow the development of efficient delayed block adaptive filters for high throughput applications.

 For the same speedup, it has been found that the selection of step size is critical for a given delay and block size, for instance, a higher delay and a lower block size lead to a slower convergence rate, and vice versa with almost the same steady-state MSE. Monte Carlo simulations for Gaussian inputs confirm the validity of the derived approximations.  

\bibliographystyle{IEEEtran}
\bibliography{IEEEabrv,ref}
%\vskip 0pt plus -1fil

%\vspace{-1cm}
	\begin{IEEEbiography}
		[{\includegraphics[width=0.85in,height=100in,clip,keepaspectratio]{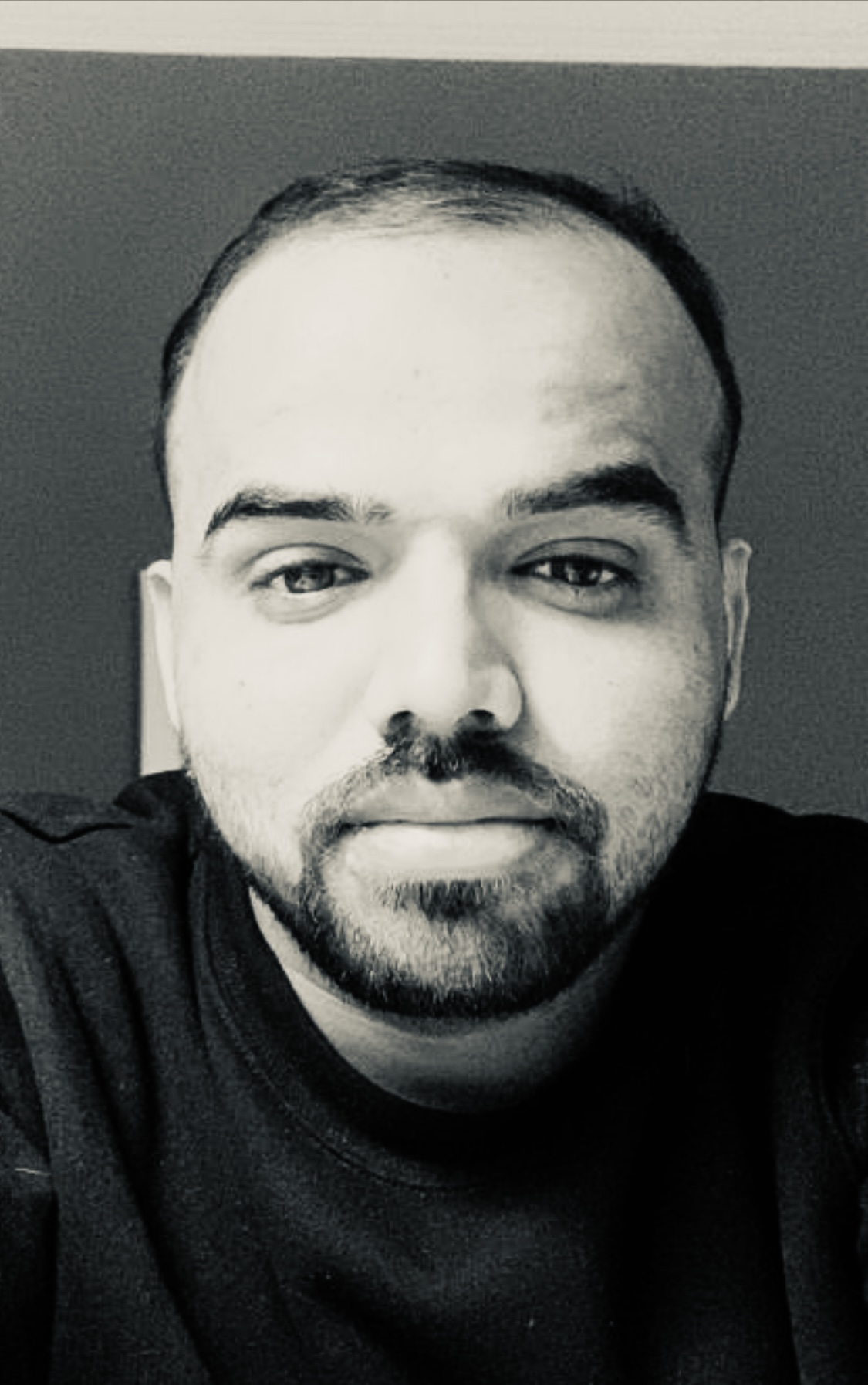}}]
		{Mohd Tasleem Khan} received the B.Tech degree in electronics, in 2013, from Zakir Hussain College of Engineering and Technology, Aligarh Muslim University, Aligarh, India. He received the Ph.D. degree in VLSI, in 2019, from Indian Institute of Technology Guwahati, India. He was a Principal Engineer at Taiwan Semiconductor Manufacturing Company (TSMC), Hsinchu, Taiwan. 
		He worked as an Assistant Professor with the Department of Electronics Engineering, Indian Institute of Technology Dhanbad, India. He is currently working as a Postdoctoral Researcher at Linköping University, Sweden. His research and teaching interests include the VLSI implementation of algorithms and architectures for signal processing, communication and machine learning applications.
	\end{IEEEbiography}
	%\vskip 0pt plus -1fil
%	\vskip 0pt plus -1fil	
	\begin{IEEEbiography}
		[{\includegraphics[width=1in,height=125in,clip,keepaspectratio]{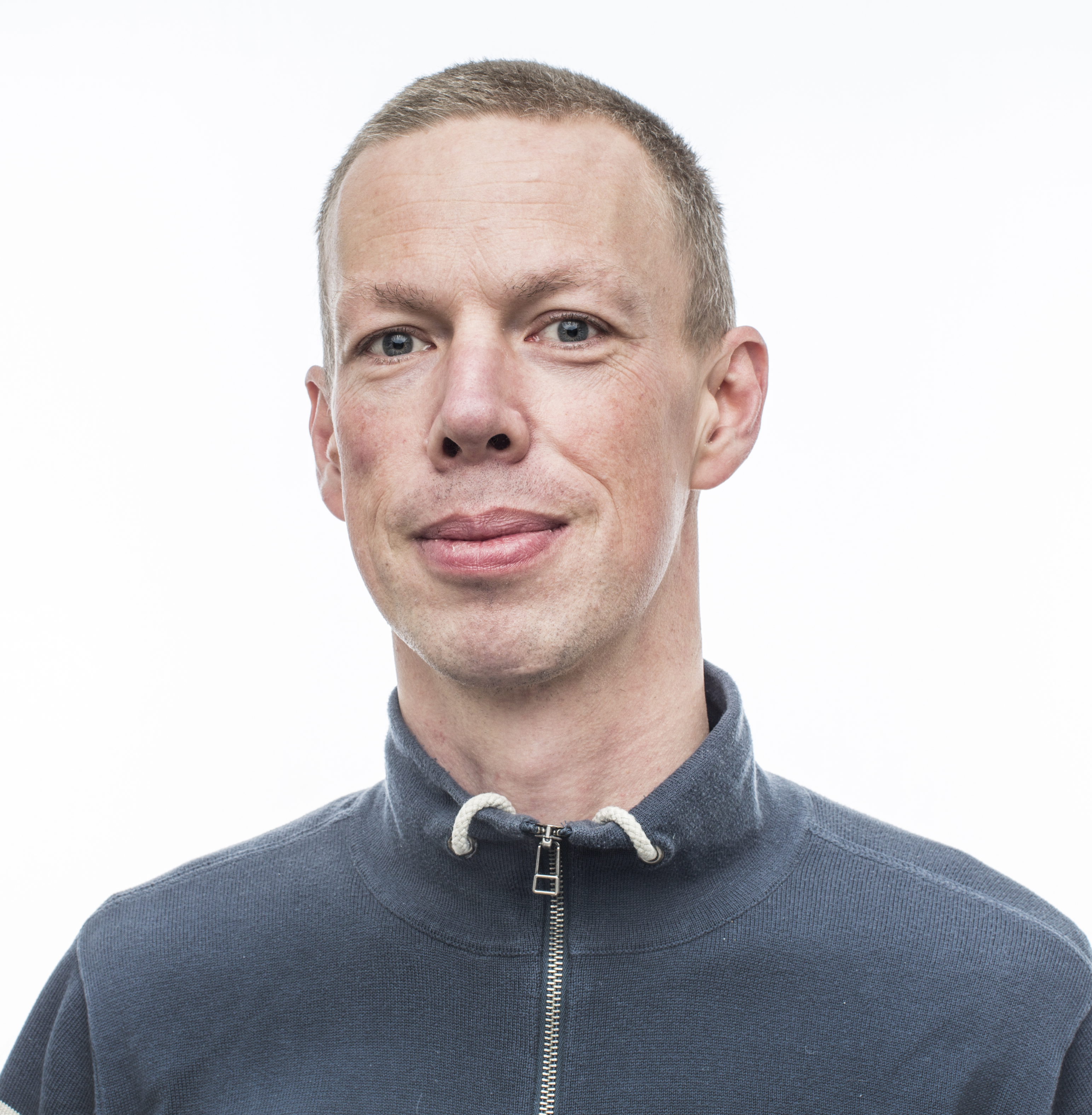}}]
		{Oscar Gustafsson} received 
		the M.Sc., Ph.D., and Docent degrees from
		Linköping University, Linköping, Sweden, in 1998,
		2003, and 2008, respectively. He is currently an Associate Professor and head of the Division of Computer Engineering, Department of Electrical Engineering at the same university.
		
		His research interests include design and implementation of algorithms and arithmetic circuits. Dr. Gustafsson has published close to 200 papers in international journal and conferences on these topics.
	\end{IEEEbiography}

	\appendices
	\numberwithin{equation}{section}
	\section{Derivation of (\ref{eq:dblms_mse_expand})}
	\label{Appendix:A}	
	By changing iteration $k$ to $k-1$ and transposing both sides of (\ref{eq:dblms_transf_simp}), we get, with $k_D = k-D-1$ for brevity:
			\begin{equation}\label{eq:A0}
	\tilde{\mathbf{v}}^T_{k}=\tilde{\mathbf{v}}^T_{k-1}-\hat\mu\tilde{\mathbf{v}}^T_{k_D}\tilde{\mathbf{X}}^T_{k_D}\tilde{\mathbf{X}}_{k_D}+\hat\mu {\mathbf{e}}^T_{\mathrm{o},k_D}\tilde{\mathbf{X}}_{k_D}+\hat\mu {\mathbf{z}}^T_{k_D}\tilde{\mathbf{X}}_{k_D}
	\end{equation}
		Taking the expectation of the composite product $\tilde{{\mathbf{v}}}^T_{k}{\mathbf{\Lambda}}\tilde{{\mathbf{v}}}_{k}$ and expand it using (\ref{eq:dblms_transf_simp}) and (\ref{eq:A0}), we get
		\begin{IEEEeqnarray}{rCl}
			\E\left[\tilde{{\mathbf{v}}}^T_{k}{\mathbf{\Lambda}}\tilde{{\mathbf{v}}}_{k}\right]& = &\E\left[\tilde{{\mathbf{v}}}^T_{k-1}{\mathbf{\Lambda}}\tilde{{\mathbf{v}}}_{k-1}\right] \nonumber \\  
			&&-\hat{\mu}\E\left[\tilde{{\mathbf{v}}}^T_{k-1}{\mathbf{\Lambda}}{\tilde{{\mathbf{X}}}^T_{k_D}}{\tilde{{\mathbf{X}}}_{k_D}}\tilde{{\mathbf{v}}}_{k_D}\right]\nonumber \\
			&&+\hat{\mu}\E\left[\tilde{\mathbf{v}}^T_{k-1}{\mathbf{\Lambda}}\tilde{\mathbf{X}}^T_{k_D}\tilde{\mathbf{e}}_{\mathrm{o},k_D}\right]\nonumber\\
			&&-\hat{\mu}\E\left[\tilde{\mathbf{v}}^T_{k_D}\tilde{\mathbf{X}}^T_{k_D}\tilde{\mathbf{X}}_{k_D}{\mathbf{\Lambda}}\tilde{\mathbf{v}}_{k-1}\right]\nonumber\\
			&&-\hat{\mu}^2\E\left[\tilde{{\mathbf{e}}}_{\mathrm{o},k_D}{\tilde{{\mathbf{v}}}^T_{k_D}}{\tilde{{\mathbf{X}}}^T_{k_D}}{\tilde{{\mathbf{X}}}_{k_D}}{\mathbf{\Lambda}}{\tilde{{\mathbf{X}}}^T_{k_D}}\right]\nonumber\\
			&&+\hat{\mu}\E\left[\tilde{\mathbf{e}}^T_{\mathrm{o},k_D}\tilde{\mathbf{X}}_{k_D}{\mathbf{\Lambda}}\tilde{\mathbf{v}}_{k-1}\right]\nonumber\\
			&&-\hat{\mu}^2\E\left[\tilde{\mathbf{e}}^T_{\mathrm{o},k_D}\tilde{\mathbf{X}}_{k_D}{\mathbf{\Lambda}}\tilde{\mathbf{X}}^T_{k_D}\tilde{\mathbf{X}}_{k_D}\tilde{\mathbf{v}}^T_{k_D}\right]\nonumber\\
			&&+\hat{\mu}^2\E\left[\tilde{{\mathbf{v}}}^T_{k_D}{\tilde{{\mathbf{X}}}^T_{k_D}}{\tilde{{\mathbf{X}}}_{k_D}}{\mathbf{\Lambda}}{\tilde{{\mathbf{X}}}^T_{k_D}}{\tilde{{\mathbf{X}}}_{k_D}}\tilde{{\mathbf{v}}}_{k_D}\right]\nonumber\\
			&&+\hat{\mu}^2\E\left[{\mathbf{e}}^T_{{\mathrm{o}},k_D}{\tilde{{\mathbf{X}}}_{k_D}}{\mathbf{\Lambda}}{\tilde{{\mathbf{X}}}^T_{k_D}}{\mathbf{e}}_{{\mathrm{o}},k_D}\right]\nonumber\\
			&&+\hat{\mu}^2\E\left[{\mathbf{z}}^T_{k_D}{\tilde{{\mathbf{X}}}_{k_D}}{\mathbf{\Lambda}}{\tilde{{\mathbf{X}}}^T_{k_D}}{\mathbf{z}}_{k_D}\right]. \label{eq:A3}
		\end{IEEEeqnarray}
		The terms containing the expectations of ${\mathbf{z}}_{k_D}$ were suppressed in (\ref{eq:A3}), except the one with ${\mathbf{z}}_{k_D}$ and ${\mathbf{z}}^T_{k_D}$. Since $\E\left[{\mathbf{e}}_{{\mathrm{o}},k_D}\right]=\E\left[{\mathbf{d}}_{k_D}]-\E[{\mathbf{X}}_{k_D}{\mathbf{w}}^{*}\right]=0$, expectations involving the first-order ${\mathbf{e}}_{{\mathrm{o}},k_D}$ term become zero. Thus,
		\begin{IEEEeqnarray}{rCl}
			\E\left[\tilde{{\mathbf{v}}}^T_{k}{\mathbf{\Lambda}}\tilde{{\mathbf{v}}}_{k}\right]& = &\E\left[\tilde{{\mathbf{v}}}^T_{k-1}{\mathbf{\Lambda}}\tilde{{\mathbf{v}}}_{k-1}\right] -\hat{\mu}\E\left[\tilde{{\mathbf{v}}}^T_{k-1}{\mathbf{\Lambda}}^2\tilde{{\mathbf{v}}}_{k_D}\right]\nonumber \\
			&&-\hat{\mu}\E\left[\tilde{{\mathbf{v}}}^T_{k_D}{\mathbf{\Lambda}}^2\tilde{{\mathbf{v}}}_{k-1}\right]\nonumber\\
			&&+\hat{\mu}^2\E\left[\tilde{{\mathbf{v}}}^T_{k_D}{\tilde{{\mathbf{X}}}^T_{k_D}}{\tilde{{\mathbf{X}}}_{k_D}}{\mathbf{\Lambda}}{\tilde{{\mathbf{X}}}^T_{k_D}}{\tilde{{\mathbf{X}}}_{k_D}}\tilde{{\mathbf{v}}}_{k_D}\right]\nonumber\\
			&&+\hat{\mu}^2\E\left[{\mathbf{e}}^T_{{\mathrm{o}},k_D}{\tilde{{\mathbf{X}}}_{k_D}}{\mathbf{\Lambda}}{\tilde{{\mathbf{X}}}^T_{k_D}}{\mathbf{e}}_{{\mathrm{o}},k_D}\right]\nonumber\\
&&+\hat{\mu}^2\E\left[{\mathbf{z}}^T_{k_D}{\tilde{{\mathbf{X}}}_{k_D}}{\mathbf{\Lambda}}{\tilde{{\mathbf{X}}}^T_{k_D}}{\mathbf{z}}_{k_D}\right].\label{eq:A4}
		\end{IEEEeqnarray}
		The number of terms in (\ref{eq:A4}) is reduced to six, whose step-by-step simplification are discussed in (\ref{eq:dblms_mse_expand})$-$(\ref{eq:dblms_mse_expand_2nd_4th}).		
		\section{Derivation of (\ref{eq:dblms_mse_expand_2nd_4th})}
		\label{Appendix:B}
		Consider an example of a system for any $L$ with delay $D=3$. The set of equations can describe (\ref{eq:dblms_mse_expand_simp_ana}) as below:
			\begin{IEEEeqnarray}{rll}
			\IEEEyesnumber\label{eq:both} \IEEEyessubnumber*
		s=1\Rightarrow\xi_{k-1,01} = & \hspace{0.1cm} \xi_{k-2,00}-\hat{\hat{\mu}} L\xi_{k-2,03}, \label{eq:Ba}\\
			s=2\Rightarrow\xi_{k-1,02} = & \hspace{0.1cm} \xi_{k-2,01}-\hat{\hat{\mu}} L\xi_{k-3,02},  \label{eq:Bb}\\
s=3\Rightarrow\xi_{k-1,03} = & \hspace{0.1cm} \xi_{k-2,02}-\hat{\hat{\mu}} L\xi_{k-4,01}.  \label{eq:Bc}
		\end{IEEEeqnarray}
	%	By changing iteration $k-1$ to $k-3$ in (\ref{eq:Ba}) and $k-1$ to $k-2$ in (\ref{eq:Bb}), we have 	
	%\begin{IEEEeqnarray}{rll}
	%	\IEEEyesnumber\label{eq:both1} \IEEEyessubnumber*
	%	s=1\Rightarrow\xi_{k-3,01} = & \hspace{0.1cm} \xi_{k-4,00}-\hat{\hat{\mu}} L\xi_{k-4,03}, \label{eq:Ba1}\\
	%	s=2\Rightarrow\xi_{k-2,02} = & \hspace{0.1cm} \xi_{k-3,01}-\hat{\hat{\mu}} L\xi_{k-4,02}.  \label{eq:Bb1}
	%\end{IEEEeqnarray}

A useful interpretation for the second terms of right hand side of (\ref{eq:Ba}) and (\ref{eq:Bb}) can be obtained from (\ref{eq:sub3}) as
%, if there is a $\delta$ change in iteration $k$ is required. This can be obtained by combining (\ref{eq:sub1}) and (\ref{eq:sub3}) for $s>\delta$ as
\begin{equation}\label{eq:interpretation1}
\xi_{k-s-1,0(r-s+1)}=\xi_{k-(s+\delta)-1,0(r-s+1-\delta)},
\end{equation}
%and, for $s\leq \delta$ as
%\begin{equation}\label{eq:interpretation2}
%\xi_{k-\delta-1,0(r-s+1)}=	\hspace{0.1cm} \xi_{k-(\delta-s+1)-1,0(r-s+1)}.
%\end{equation}
%Clearly, (\ref{eq:Ba1}) and (\ref{eq:Bb1}) satisfy $s>\delta$, therefore they can be further reduced to  
Using (\ref{eq:interpretation1}), we can simplify (\ref{eq:Ba}) and (\ref{eq:Bb}) as
\begin{IEEEeqnarray}{rll}
	\IEEEyesnumber\label{eq:both2} \IEEEyessubnumber*
	s=1\Rightarrow\xi_{k-1,01} = & \hspace{0.1cm} \xi_{k-2,00}-\hat{\hat{\mu}} L\xi_{k-4,01}, \label{eq:Ba2}\\
	s=2\Rightarrow\xi_{k-1,02} = & \hspace{0.1cm} \xi_{k-2,01}-\hat{\hat{\mu}} L\xi_{k-4,01}.  \label{eq:Bb2}
\end{IEEEeqnarray}
%To further simplify (\ref{eq:both2}), we need to transform (\ref{eq:interpretation1}) as 
%\begin{equation}\label{eq:interpretation3}
%\xi_{k-(s-\delta)-1,0(r-s+1)} = \xi_{k-(s-\delta)-\gamma-1,01},
%\end{equation}
The terms at the left hand side of (\ref{eq:Ba2}) and (\ref{eq:Bb2}) are substituted in (\ref{eq:Bb2}) and (\ref{eq:Bc}), respectively. This requires unit delay versions of the terms at the left-hand side of (\ref{eq:Ba2}) and (\ref{eq:Bb2}) using the following interpretation based on combining (\ref{eq:sub1}) and (\ref{eq:sub3}) as   
\begin{equation}\label{eq:interpretation2}
\xi_{k-s-1-\gamma,0(r-s+1)}=\xi_{k-(s-\gamma)-1,0(r-s+1)},
\end{equation}
From this, $\xi_{k-1,02}=\xi_{k-2,02}$ and $\xi_{k-1,01}=\xi_{k-2,01}$. Finally, substituting (\ref{eq:Ba2}) and (\ref{eq:Bb2}) into (\ref{eq:Bb2}) and (\ref{eq:Bc}), respectively, we have 
%where $\gamma=r-s$. Using (\ref{eq:interpretation3}), one can deduce $\xi_{k-2,03}=\xi_{k-4,01}$ and $\xi_{k-3,02}=\xi_{k-4,01}$, and combining (\ref{eq:Bc}), (\ref{eq:Ba2}) and (\ref{eq:Bb2}), we establish the following relation:		
\begin{equation}\label{eq:Bd}
\xi_{k-1,03}=\xi_{k-4,00}-3\hat{\hat{\mu}} L\xi_{k-4,01}.
\end{equation}
In general, (\ref{eq:Bd}) can be extended for any $s=D$ as required in (\ref{eq:dblms_mse_expand_2nd_4th}) to get $\xi_{k-1,0D}$ as
\begin{equation}\label{eq: Bf}
\xi_{k-1,0D}=\xi_{k_D,00}-\hat{\hat{\mu}} LD\xi_{k_D,01}.
\end{equation}		 
So, (\ref{eq: Bf}) holds for any $D$ under first order approximations.  

		\section{Derivation of (\ref{eq:D3})}
		\label{Appendix:C}
		Unlike (\ref{eq:A3}), we first need to expand 
		$\tilde{\mathbf{v}}^T_{k}$ using (\ref{eq:A0}) to compute $\tilde{\mathbf{v}}^T_{k}\mathbf{\Lambda}\tilde{\mathbf{v}}_{k}$ as
		\begin{IEEEeqnarray}{rCl}\label{eq:C1}
			\tilde{\mathbf{v}}^T_{k}\mathbf{\Lambda}\tilde{\mathbf{v}}_{k}&=&\tilde{\mathbf{v}}^T_{k-1}\mathbf{\Lambda}\tilde{\mathbf{v}}_{k}+\hat\mu {\mathbf{e}}^T_{\mathrm{o},k_D}\tilde{\mathbf{X}}_{k_D}\mathbf{\Lambda}\tilde{\mathbf{v}}_{k}\nonumber\\ 
			&& -\hat\mu\tilde{\mathbf{v}}^T_{k_D}\tilde{\mathbf{X}}^T_{k_D}\tilde{\mathbf{X}}_{k_D}\mathbf{\Lambda}\tilde{\mathbf{v}}_{k}.
		\end{IEEEeqnarray}
		Then, we substitute $\tilde{\mathbf{v}}_{k}$ from (\ref{eq:dblms_transf_simp}) by changing the iteration $k$ to $k-1$ in first term of (\ref{eq:C1}), to get
			\begin{IEEEeqnarray}{rCl}\label{eq:C2}
		\tilde{\mathbf{v}}^T_{k}\mathbf{\Lambda}\tilde{\mathbf{v}}_{k}&=&\tilde{\mathbf{v}}^T_{k-1}\mathbf{\Lambda}\tilde{\mathbf{v}}_{k-1}+{\hat\mu}\tilde{\mathbf{v}}^T_{k-1}\mathbf{\Lambda}\tilde{\mathbf{X}}^T_{k_D}{\mathbf{e}}_{\mathrm{o},k_D}
		\nonumber\\
		&&-{\hat\mu}\tilde{\mathbf{v}}^T_{k-1}\mathbf{\Lambda}\tilde{\mathbf{X}}^T_{k_D}\tilde{\mathbf{X}}_{k_D}\tilde{\mathbf{v}}_{k_D}
		 +\hat\mu {\mathbf{e}}^T_{\mathrm{o},k_D}\tilde{\mathbf{X}}_{k_D}\mathbf{\Lambda}\tilde{\mathbf{v}}_{k}\nonumber\\
		&& -\hat\mu\tilde{\mathbf{v}}^T_{k_D}\tilde{\mathbf{X}}^T_{k_D}\tilde{\mathbf{X}}_{k_D}\mathbf{\Lambda}\tilde{\mathbf{v}}_{k}.
	\end{IEEEeqnarray}
Taking expectation of (\ref{eq:C2}), and keeping  $\E\left[{\mathbf{e}}_{{\mathrm{o}},k_D}\right]=\E\left[{\mathbf{d}}_{k_D}]-\E[{\mathbf{X}}_{k_D}{\mathbf{w}}^{*}\right]=0$, we obtain
		\begin{IEEEeqnarray}{rCl}\label{eq:C4}
			\E\left[\tilde{\mathbf{v}}^T_{k}\mathbf{\Lambda}\tilde{\mathbf{v}}_{k}\right]&=&\E\left[\tilde{\mathbf{v}}^T_{k-1}\mathbf{\Lambda}\tilde{\mathbf{v}}_{k-1}\right]\nonumber \\
			&&-\hat{\hat\mu}\E\left[\tilde{\mathbf{v}}^T_{k-1}\mathbf{\Lambda}\tilde{\mathbf{X}}^T_{k_D}\tilde{\mathbf{X}}_{k_D}\tilde{\mathbf{v}}_{k_D}\right]\nonumber \\
			&& -\hat{\hat\mu}\E\left[\tilde{\mathbf{v}}^T_{k_D}\tilde{\mathbf{X}}^T_{k_D}\tilde{\mathbf{X}}_{k_D}\mathbf{\Lambda}\tilde{\mathbf{v}}_{k-1}\right].
		\end{IEEEeqnarray}
		\begin{IEEEeqnarray}{rCl}\label{eq:C5}
	\E\left[\tilde{\mathbf{v}}^T_{k}\mathbf{\Lambda}\tilde{\mathbf{v}}_{k}\right]&=&\E\left[\tilde{\mathbf{v}}^T_{k-1}\mathbf{\Lambda}\tilde{\mathbf{v}}_{k-1}\right]\nonumber \\
	&&-2\hat{\hat\mu}\E\left[\tilde{\mathbf{v}}^T_{k-1}\mathbf{\Lambda}\tilde{\mathbf{X}}^T_{k_D}\tilde{\mathbf{X}}_{k_D}\tilde{\mathbf{v}}_{k_D}\right].
\end{IEEEeqnarray}
Unlike (\ref{eq:A4}), the number of terms in (\ref{eq:C5}) is two. 
		
\section{Derivation of (\ref{eq:shorthand_matrix_vector})}		\label{Appendix:D}	
By changing $k$ to $k-2$ in (\ref{eq:Ba}); $k$ to $k-1$ in (\ref{eq:Bb}) and re-arranging (\ref{eq:Ba}), (\ref{eq:Bb}) and (\ref{eq:Bc}), we have
\begin{IEEEeqnarray}{rCl} \label{eq:Da}	 
	\xi_{k-4,00}&=&\xi_{k-3,01}+\hat{\hat{\mu}} L\xi_{k-4,03} \nonumber\\
	0&=&-\xi_{k-3,01}+\xi_{k-2,02}+\hat{\hat{\mu}} L\xi_{k-4,02} \nonumber\\
	0&=&\hat{\hat{\mu}} L\xi_{k-4,01}-\xi_{k-2,02}+\xi_{k-1,03}.  
\end{IEEEeqnarray}
Using (\ref{eq:interpretation2}), we simplify (\ref{eq:Da}) as
\begin{IEEEeqnarray}{rCl} \label{eq:Db}	 
	\xi_{k-4,00}&=&\xi_{k-4,01}+\hat{\hat{\mu}} L\xi_{k-4,03} \nonumber\\
	0&=&-\xi_{k-4,01}+\xi_{k-4,02}+\hat{\hat{\mu}} L\xi_{k-4,02} \nonumber\\
	0&=&\hat{\hat{\mu}} L\xi_{k-4,01}-\xi_{k-4,02}+\xi_{k-4,03}  .
\end{IEEEeqnarray}
Equation~(\ref{eq:Db}) can be written in matrix-vector multiplication form as
\begin{equation}\label{eq:Dc}
\begin{bmatrix}
\xi_{k-4,00} \\ 
0 \\
0   
\end{bmatrix}=
\begin{bmatrix}
1 & 0 & \hat{\hat{\mu}} L\\ 
-1 & 1+\hat{\hat{\mu}} L & 0 \\
\hat{\hat{\mu}} L & -1 & 1
\end{bmatrix}
\begin{bmatrix}
 \xi_{k-4,01}\\ 
  \xi_{k-4,02} \\  
 \xi_{k-4,03}
\end{bmatrix}.
\end{equation}	
Similarly, the analysis for $D=4$ can be extended for any $L$. The corresponding matrix-vector form is then obtained as:
\begin{equation}\label{eq:Dd}
\begin{bmatrix}
\xi_{k-5,00} \\ 
0 \\
0    \\
0
\end{bmatrix}=
\begin{bmatrix}
1 & 0 & 0 & \hat{\hat{\mu}} L\\ 
-1 & 1 & \hat{\hat{\mu}} L & 0 \\
0 & -1+\hat{\hat{\mu}} L & 1 & 0 \\
\hat{\hat{\mu}} L & 0 & -1 & 1
\end{bmatrix}
\begin{bmatrix}
\xi_{k-5,01}\\ 
\xi_{k-5,02} \\  
\xi_{k-5,03} \\
\xi_{k-5,04} \\
\end{bmatrix}.
\end{equation}	
Like (\ref{eq:Dc}) and (\ref{eq:Dd}), the matrix evolves in the same fashion for odd $D$ and even $D$ respectively. However, in general, the matrix can be described as:
\begin{equation}\label{eq:De}
\begin{bmatrix}
\xi_{k_D,00} \\
0 \\ 
\vdots \\
0 
\end{bmatrix}=
\begin{bmatrix}
1 & 0 & \cdots & \hat{\hat{\mu}} L\\
-1 & 1 & \cdots & 0\\ 
\vdots & \vdots & \ddots & \vdots\\ 
\hat{\hat{\mu}} L & 0 & \cdots & 1 
\end{bmatrix}
\begin{bmatrix}
\xi_{k_D,01}\\
\xi_{k_D,02}\\ 
\vdots\\ 
\xi_{k_D,0D} 
\end{bmatrix}.
\end{equation}

In order to determine $\alpha_{\hat{\hat\mu} L,D}$ as given in (\ref{eq: dev factor}), it is required to find $\left|{\mathbf{S}}_{\hat{\hat\mu} L,D}\right|$ and $\left|{\mathbf{S}}_{-\hat{\hat\mu} L,(D-1)}\right|$ for any $D$. Consider $\left|{\mathbf{S}}_{\hat{\hat\mu} L,D}\right|$ with $D=3$ using (\ref{eq:Dc}) as
\begin{IEEEeqnarray}{rCl}
	\left|{\mathbf{S}}_{\hat{\hat\mu} L,3}\right|&= & \begin{vmatrix}
		1+\hat{\hat\mu} L & 0\\
		-1 & 1 \\
	\end{vmatrix}+\hat{\hat\mu} L\begin{vmatrix}
		-1 & 1+\hat{\hat\mu} L\\
		\hat{\hat\mu} L & -1 \\
	\end{vmatrix} \nonumber\\
	&= & \left|{\mathbf{S}}_{\hat{\hat\mu} L,1}\right|+\hat{\hat\mu} L\left|{\mathbf{S}}_{-\hat{\hat\mu} L,2}\right|.
\end{IEEEeqnarray}
In general, we obtain the following recursive rule for $|{\mathbf{S}}_{\hat{\hat\mu} L,D}|$ for any $D$ and $L$ as  
\begin{equation}\label{eq:det2}
\left|{\mathbf{S}}_{\hat{\hat\mu} L,D}\right|=\left|{\mathbf{S}}_{\hat{\hat\mu} L,D-2}\right|
+\hat{\hat\mu} L\left|{\mathbf{S}}_{-\hat{\hat\mu} L,D-1}\right|, \ D\geq 2,
\end{equation}	
with $\left|{\mathbf{S}}_{\hat{\hat\mu} L,0}\right|=1$ and $\left|{\mathbf{S}}_{\hat{\hat\mu} L,1}\right|=1+\hat{\hat\mu} L$. The above recursive rule can be proven by induction. From this, $\left|{\mathbf{S}}_{\hat{\hat\mu} L,3}\right|$ can be expressed explicitly as 
\begin{IEEEeqnarray}{rCl}\label{eq:det_3}
	\left|{\mathbf{S}}_{\hat{\hat\mu} L,3}\right|&=&\left|{\mathbf{S}}_{\hat{\hat\mu} L,1}\right|
	+\hat{\hat\mu} L\left|{\mathbf{S}}_{-\hat{\hat\mu} L,2}\right|\nonumber \\
	&=&1+2\hat{\hat\mu} L -\left(\hat{\hat\mu} L\right)^2-\left(\hat{\hat\mu} L\right)^3.
\end{IEEEeqnarray}

A similar analysis for $\left|{\mathbf{S}}_{\hat{\hat\mu} L,D}\right|$ with $D=4$ using (\ref{eq:Dd}) can also be carried out. Thus, a general explicit form of $\left|{\mathbf{S}}_{\hat{\hat\mu} L,D}\right|$ for any $D$ and $L$ is given in (\ref{eq:det_general}).

		\end{document}